\documentclass[aps,pra,showpacs,twoside,10pt,floatfix,nofootinbib,longbibliography,twocolumn]{revtex4-2}
\usepackage[colorlinks=true, citecolor=red, urlcolor=blue ]{hyperref}
\usepackage{epsfig,newlfont,amssymb,amsfonts,amsmath,bm,subfigure,palatino,mathtools,amsthm,braket,soul,enumitem,color,times,comment,geometry,xcolor,contour,enumitem,mathrsfs}
\usepackage[normalem]{ulem}
\begin{document}

\title{A graph-aware bounded distance decoder for all stabilizer codes}
\author{Harikrishnan K J and Amit Kumar Pal}
\affiliation{Department of Physics, Indian Institute of Technology Palakkad, Palakkad 678 623, India}
\date{\today}

\begin{abstract}
We formulate a bounded distance decoding strategy applicable to all stabilizer codes including both CSS and non-CSS code-families. The framework emerges out of the local Clifford equivalence between arbitrary stabilizer states and graph states. Using the graphical representation of the stabilizers and the syndromes, we constitute the bounded distance decoding as an adaptable generalization of maximum likelihood decoding, ensuring correction of all errors with weights upper bounded by a target weight. We show that strategic pruning associated with a feed-forward network structure of the graph can reduce the search space and subsequently the runtime of the designed decoder.  We demonstrate satisfactory performance of the bounded distance decoder in the case of the optimal non-CSS codes up to distance $d=11$ subjected to the depolarizing error on all qubits, and near-optimal decoding for the color and the surface codes, both belonging to the CSS family, under the bit-flip errors on the qubits. We also develop an open-source library, \href{https://github.com/harikrishnan9779/QGDecoder}{QGDecoder}, enabling the graph-aware bounded distance decoding of arbitrary stabilizer codes. 
\end{abstract}

\maketitle

\section{Introduction}

The landscape of quantum technology is rapidly advancing through increasingly large-scale and utility-driven experiments~\cite{Acharya2025,Abanin2025,Kim2023}, enabling the transition from noisy intermediate-scale quantum (NISQ)~\cite{Preskill2018,Cheng2023,Yan2025} to fault-tolerant application-scale quantum~\cite{Eisert2025} devices. Starting from Shor’s pioneering work~\cite{Shor1995}, quantum error correction (QEC)~\cite{Gottesman2009,Nielsen2010,Terhal2015,Roffe2019,Girvin2023} remains central to this progress, attracting extensive efforts on understanding fault-tolerance and developing improved QEC codes~\cite{Knill1997,Knill1998,Eastin2009,Gottesman1997,Laflamme1996,Calderbank1996,Steane1996,Kitaev2003,Bombin2006,Tillich2009}.  Out of the ever-widening catalog~\cite{ErrorCorrectionZoo,Grassl_codetables} of stabilizer codes~\cite{Gottesman1997}, two broad classes emerge based on their structural properties, namely, the Calderbank–Shor–Steane (CSS) codes, and the non-CSS codes. Among the former,    the surface~\cite{Kitaev2003,Fowler2012} and the color~\cite{Bombin2006,Landahl2011} codes, belonging to the low-density parity-check (LDPC)~\cite{Breuckmann2021,Tillich2009} category, have been identified as leading candidates, particularly due to their suitability for the planar architecture of current superconducting quantum processors~\cite{Fowler2012,Acharya2025} as well as the availability of inherently fault-tolerant transversal controlled-NOT gate~\cite{Gottesman1997}. Parallely, non-CSS codes have been under focus due to their observed efficiency in terms of offering a higher number of logical qubits and a longer distance for a fewer number of physical qubits over the CSS codes~\cite{Laflamme1996,Calderbank1996,Steane1996}, particularly in the current era of NISQ processors built on around a hundred physical qubits~\cite{Acharya2025,Kim2023}. This is further highlighted by the construction of the XZZX code belonging to the non-CSS family, which is found to perform remarkably well against biased noises~\cite{Bonilla2021}. 

Even for a well-designed code, keeping the code in the logical space and performing computations with the logical qubits, real-time decoding~\cite{Terhal2015,Battistel2023,Acharya2025} is unavoidable, but creates a bottleneck~\cite{deMarti2024,Gottesman2009}. Optimal decoding considers degeneracy of errors in quantum codes, and is a \#P-complete problem for all stabilizer codes~\cite{Iyer2015,Hsieh2011}. There also exist suboptimal non-degenerate strategies, referred to as maximum likelihood decoding (MLD)~\cite{MacWilliams1977,Iyer2015,deMarti2024}, that are NP-complete~\cite{Iyer2015,Hsieh2011},  completely disregards the effects of degeneracy, and obtain the minimum weight error configuration conforming to the error syndrome under identical independently distributed noise models~\cite{Iyer2015,deMarti2024}. 

Albeit both decoding strategies are computationally challenging for large distance generic codes, there exist efficient algorithms for specific code families, such as the CSS codes~\cite{Bravyi2014,Fowler2014,Higgott2022,Delfosse2022,Wu2023,Sabo2024,deMarti2024,Higgott2025}. For example, surface codes are decoded with tensor network decoders~\cite{Bravyi2014},  matching decoders~\cite{Fowler2014,Higgott2025,Wu2023}, and union-find decoders~\cite{Delfosse2021} (see \cite{deMarti2024} for a comparison in terms of decoding optimality and efficiency), while for color codes, some specialized variants of matching algorithms~\cite{Lee2025} as well as a MaxSAT formulation~\cite{Berent2024} exist. Further, belief propagation algorithms~\cite{Pearl1988,Kschischang2001,Sae2001} equipped with suitable post-processing methods~\cite{Poulin2008,Panteleev2021,Hillmann2025} have become a useful tool in exploring CSS LDPC codes~\cite{Roffe2020}.

Extending these decoders to non-CSS codes is generally non-trivial, owing to a range of underlying challenges. A variant of the matching algorithm has been developed for the XZZX code, but it leverages the specific geometric structure of the code and is therefore not readily generalizable~\cite{Bonilla2021}. A belief propagation strategy can also, in principle, be formulated for arbitrary stabilizer codes~\cite{Poulin2008}, as it relies primarily on the pattern of nontrivial actions of stabilizers on the qubits which is encoded in a Tanner graph~\cite{Tanner1981}, together with the noise model parameters~\cite{Poulin2008, Panteleev2021}. However, when applied to non-CSS codes, particularly under depolarizing noise, belief propagation may, in addition to producing logical errors, fail to converge to any error consistent with the observed syndrome. Such failures are typically treated as decoding failures~\cite{Poulin2008}. For this reason, the word error rate is often used as a more appropriate performance metric than the logical error rate~\cite{Panteleev2021}. Given these limitations despite the clear motivation for investigating non-CSS codes, in this paper, we ask whether it is possible to devise a generic decoding framework for arbitrary stabilizer codes that never results in a decoder failure when applied to non-CSS codes. We answer this question affirmatively by adopting a perspective distinct from the conventional stabilizer-code decoding literature~\cite{Higgott2022,Delfosse2022,Sabo2024,Panteleev2021,deMarti2024,Lee2025,Berent2024,Bonilla2021} by working through the framework of graph states~\cite{Hein2006}.

Graphs and corresponding graph states~\cite{Hein2006} have found wide applications across quantum information theory, including the study of genuine multipartite entanglement~\cite{Hein2004}, measurement-based quantum computation~\cite{Raussendorf2000,Raussendorf2001,Briegel2009}, and quantum many-body phenomena~\cite{Plodzien2025}. Combined with an algebraic structure, they also provide a natural framework for constructing QEC codes~\cite{Schlingemann2001,Schlingemann2001_b,schlingemann2002}, which can be further connected to cluster states~\cite{Schlingemann2001,Schlingemann2001_b,schlingemann2002,Raussendorf2000,Raussendorf2001}. Within the codeword-stabilized (CWS) formalism, a graph, together with a classical code, specifies a quantum code~\cite{Cross2009,Chuang2009,Sowrabh2024}, and any such code can be brought, via local Clifford transformations, to a standard form defined by graph-state generators~\cite{Cross2009}. Alternative graph-based approaches, such as  mapping stabilizer codes to encoder-respecting graphs~\cite{Khesin2025}, clustered decoding explicitly for CWS codes~\cite{Li2010,Li2010_b,Li2010_c}, and specialized decoders for restricted families of graph codes~\cite{Basak2026} have also been developed.  

Similar to the CWS formalism, in our framework, we exploit the fact that arbitrary stabilizer states are local Clifford equivalent to graph states~\cite{Amaro2020,VandenNest2004}, allowing an $[[N,k,d]]$ stabilizer code and the corresponding stabilizer state to be associated respectively to a graph on the $N$ nodes, and the corresponding graph state up to local Clifford operations. Using the emergent graph, we employ a bounded distance decoding (BDD) strategy~\cite{MacWilliams1977,Liu2011,Kasai2012,Aggarwal2025,Li2010_c} ensuring that errors up to a target weight are always corrected. The approach can be seen as an adaptable generalization of MLD~\cite{MacWilliams1977,Iyer2015,deMarti2024}, and is applicable to all stabilizer codes, including both CSS and non-CSS families. We explore the runtime reduction of the developed BDD via appropriate pruning of the graph laid out as a feed-forward network. We demonstrate the applicability of the BDD in a number of optimal non-CSS codes as well as the color and the surface codes from the CSS family. The decoder performs satisfactorily in the case of the optimal non-CSS codes up to distance $d=11$ subjected to the depolarizing error, and near-optimally for the color and the surface codes under the bit-flip error. We also enable the graph-aware bounded distance decoding of arbitrary stabilizer codes through an open-source library, referred to as the \href{https://github.com/harikrishnan9779/QGDecoder}{QGDecoder}.

The rest of the paper is organized as follows. In Sec.~\ref{sec:preliminaries}, we discuss the equivalent graph description for arbitrary stabilizer codes (Sec.~\ref{subsec:equivalent_graphs}), the mapping of stabilizer syndromes to the graph syndromes (Sec.~\ref{subsec:stab_to_graph_syndrome}), and the coset structure of Pauli errors (Sec.~\ref{subsec:cosets}). The proposal of graph-aware bounded distance decoding is put forward in Sec.~\ref{sec:bounded_distance}, where Sec.~\ref{subsec:bdd} contains the formal definition, followed by a discussion in Sec.~\ref{subsec:FFN_pruning_struct_sampling} on runtime reduction via graph pruning and structured sampling, and  a demonstration of the performance of the developed decoder in optimal non-CSS codes in  Sec.~\ref{subsec:optimal_codes}. The application of the decoder in the case of CSS codes is discussed in Sec.~\ref{sec:CSS}, with specific examples of the color and the surface codes (Sec.~\ref{subsec:color+surface}). Sec.~\ref{sec:conclusion} contains concluding remarks and outlook.

\section{Setting up the graph representations}
\label{sec:preliminaries}

In this section, we briefly  discuss the properties of stabilizer codes used for QEC, and the methodology for obtaining an equivalent graph from a QEC code. Also, we describe how the error syndromes can be mapped onto equivalent graph syndromes, which is required decoding error using a graphical perspective.

\begin{figure*}
    \centering
    \includegraphics[width=0.7\linewidth]{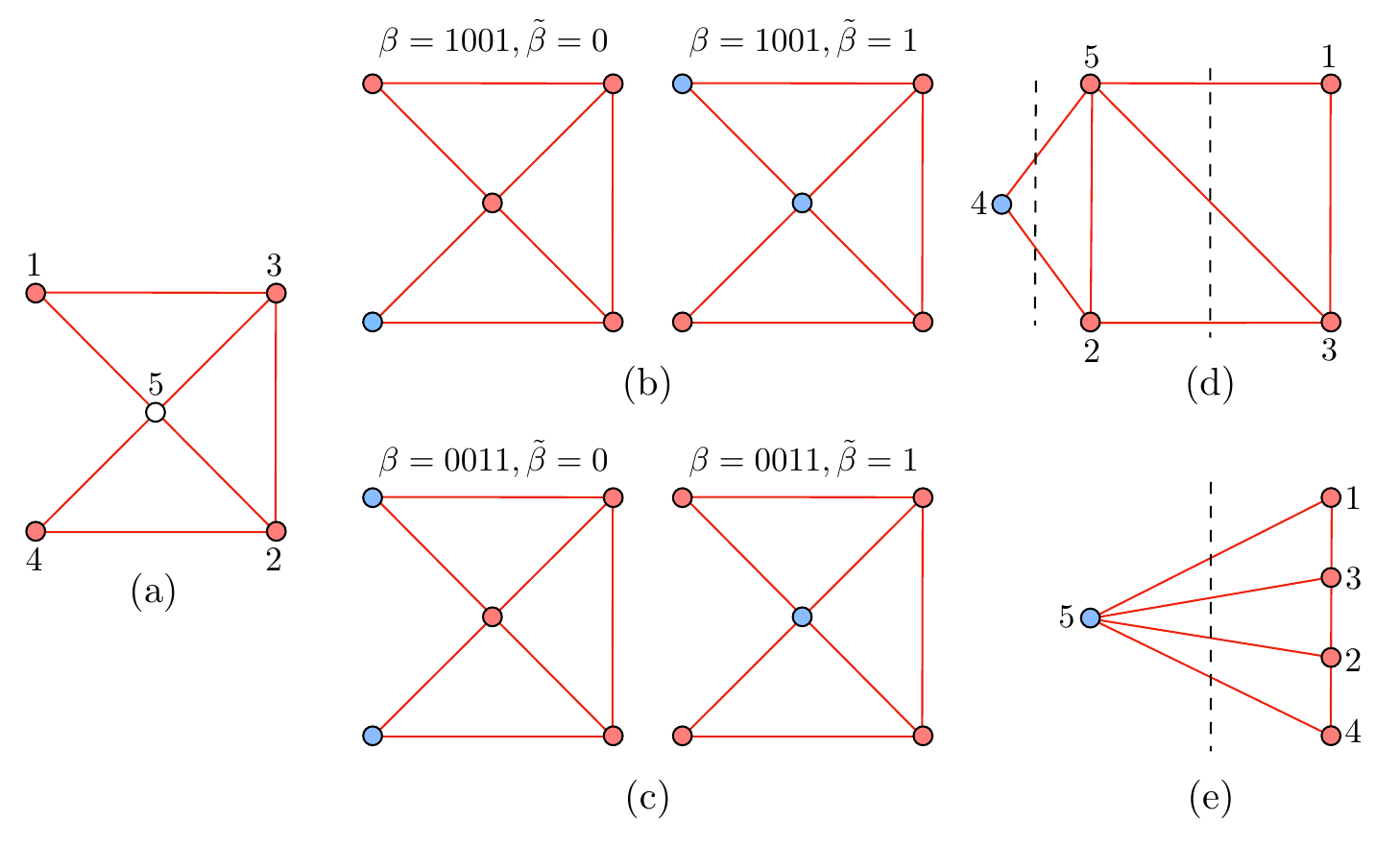}
    \caption{(a) The equivalent graph associated with the $[[5,1,3]]$ perfect code. See Appendix.~\ref{app:5_1_3_perfect_code} for the stabilizers and a detailed discussion. White node indicate the Hadamard in the $H_r$. A demonstration of resulting graph syndromes in the case of (b) $\sigma^z_4$ and (c) $\sigma^x_5$ errors where the blue nodes indicates the non-trivial graph generators of $\alpha$. The corresponding feedforward network corresponding to the minimum weight scenarios (b) $\sigma^z_4$ error where $\beta=1001,\tilde\beta=0$ and (d) $\sigma^x_5$ error where $\beta=0011,\tilde\beta=1$.}
    \label{fig:5_1_3_perfect_code}
\end{figure*}

\subsection{Equivalent graphs from stabilizer codes}
\label{subsec:equivalent_graphs}

Let us consider a stabilizer code, denoted by  $[[N,k,d]]$, defined on $N$ physical qubits that encode $k$ logical qubits. The code has a code-distance $d$, i.e., it can detect (correct) errors acting non-trivially on $2t$ $(t)$ physical qubits~\cite{Gottesman1997,Gottesman2009,Nielsen2010}, where
\begin{equation}\label{eq:t}
t=\frac{d-1}{2}.    
\end{equation}
The code space of the code is the common $+1$ eigenspace of $N-k$ linearly independent set of stabilizer operators denoted by 
\begin{equation}\label{eq:stabilizers}
    \mathcal{S}^c=\{\mathcal{S}_1^c,\mathcal{S}_2^c,\cdots,\mathcal{S}_{N-k}^c\},
\end{equation}
and is a manifold of $2^k$ \emph{code words} that are common eigenstates of the set of $k$ logical operators 
\begin{equation}\label{eq:logicals}
    \mathcal{L}^z=\{\mathcal{L}^z_1,\mathcal{L}^z_2,\cdots,\mathcal{L}^z_k\},
\end{equation}
with eigenvalues $\pm1$. The common $+1$ eigenstate of all the $k$ logical operators is a \emph{unique} stabilizer state $\ket{\mathcal{S}}$, given by 
\begin{eqnarray}\label{eq:stabilizer_state}
\mathcal{S}_j\ket{\mathcal{S}}=+1\ket{\mathcal{S}}\;\forall\; j,
\end{eqnarray}
which is \emph{stabilized} by the set of stabilizers
\begin{equation}\label{eq:total_stabilizers}
    \mathcal{S}=\mathcal{S}^c\cup \mathcal{L}^z.
\end{equation}
All $N$-qubit stabilizer states can be represented with the check matrix $A_\mathcal{S}\in \text{GF}(2)$ of size $2N\times N$ in the binary picture~\cite{Gottesman1997,Nielsen2010,Amaro2020,VandenNest2004}, with
\begin{equation}\label{eq:A_S_arbitrary}
    A_\mathcal{S}=\begin{pmatrix}
        \mathcal{Z}\\\mathcal{X}
    \end{pmatrix}.
\end{equation}
The matrix elements $\mathcal{Z}_{ij}$ and $\mathcal{X}_{ij}$ of the $N\times N$ matrices $\mathcal{Z}$ and $\mathcal{X}$ respectively specify the non-trivial action of $\mathcal{S}_j$ on the $i^{\text{th}}$ qubit (see Appendix.~\ref{app:stab_to_graph} for details).

A graph, $G$, is a collection of a set of $N$ nodes denoted by $V=\{1,2,\cdots,N\}$, and a set of links connecting the nodes, denoted by $L=\{(i,j):\forall (i,j)\in G\}$~\cite{Diestel2000}. We consider only \emph{simply connected} and \emph{undirected} graphs, where at most one link between every two nodes, and no links connecting a node to itself exist, and no specific direction is assigned to the links (see Fig.~\ref{fig:5_1_3_perfect_code}). The connectivity between the nodes of the graph is given by a real symmetric matrix, referred to as the \emph{adjacency matrix} and denoted by $\Gamma\in \text{GF}(2)$, whose matrix elements are given by
\begin{eqnarray}\label{eq:adjacency_matrix}
    \Gamma_{i,j}&=&1,\hspace{0.2cm}\forall (i,j)\in L,\nonumber\\
    \Gamma_{i,j}&=&0,\hspace{0.2cm}\forall (i,j)\notin L.
\end{eqnarray}
Each simply connected undirected graph corresponds to a graph state, $\ket{G}$, which is the common $+1$ eigenstate of $N$ stabilizers in the form of the graph generators associated to every node of the graph, defined by
\begin{equation}\label{eq:graph_generators}    g_i=\sigma^x_i\bigotimes_{j\in\mathcal{N}_i}\sigma^z_j,\;i=1,2,\cdots,N, 
\end{equation}
where $\mathcal{N}_i$ denotes the neighborhood of the $i^{\text{th}}$ node. Therefore, the check matrix corresponding to a graph state takes the form
\begin{equation}
    A_g=\begin{pmatrix}
        \Gamma\\I
    \end{pmatrix}.
\end{equation}

Since all stabilizer states are local Clifford equivalent to a graph state~\cite{Amaro2020,VandenNest2004}, it is always possible to find transformation $A_\mathcal{S}\to A_g$ for arbitrary $A_\mathcal{S}$. Since the stabilizers are linearly independent, $A_\mathcal{S}$ is of full rank $N$. Assuming $n$ to be the rank of $\mathcal{X}$, the check matrix can always be transformed to (see Appendix.~\ref{app:stab_to_graph} for details)
\begin{equation}\label{eq:A_Sp_l_r}
        A_\mathcal{S}^\prime=\begin{pmatrix}
        \mathcal{Z}_{1l} & \mathcal{Z}_{2l}\\
        \mathcal{Z}_{1r} & \mathcal{Z}_{2r}\\
        \mathcal{X}_{1l} & 0\\
        \mathcal{X}_{1r} & 0
    \end{pmatrix},
\end{equation}
via rearrangements and recombinations among the columns, where $\mathcal{X}_{1l}$ $(\mathcal{Z}_{2r})$ is an invertible $n\times n$ $(N-n\times N-n)$ sub-matrix of $\mathcal{X}$ $(\mathcal{Z})$, and the remaining sub-matrices are obtained by keeping appropriate ordering of qubits. From now onwards, we refer to the qubits belonging to $\mathcal{X}_{1l}$ $(\mathcal{Z}_{2r})$ by the left (right) nodes. The graph corresponding to $A_\mathcal{S}$ can be extracted as~\cite{Amaro2020}
\begin{equation}\label{eq:A_S_to_A_G}
    A_g=R_lH_rA_\mathcal{S}^\prime J,
\end{equation}
where $H_r$ $(R_l)$ denotes Hadamard (phase gate) operations on all of the right nodes (selected left nodes) (see Appendix.~\ref{app:stab_to_graph} for details). The stabilizer recombination matrix $J$, and finally the adjacency matrix $\Gamma$ of the associated graph $G$ can be obtained in terms of $\mathcal{X}$ and $\mathcal{Z}$ sub-matrices (see Appendix.~\ref{app:stab_to_graph} for details). 

\subsection{Graph syndroms from stabilizer syndromes}
\label{subsec:stab_to_graph_syndrome}

Any correctable Pauli error can be detected by measuring the $N-k$ stabilizers, denoted by $\mathcal{S}_j^c$ (Eq.~(\ref{eq:stabilizers})) with the superscript ``$c$" denoting the code space,  resulting in outcomes $\pm1$. We call the $(N-k)$-bit string $\beta\in\text{GF}(2)$, with
\begin{equation}\label{eq:stab_syndrome}
\beta=\beta_1\beta_2\cdots\beta_{N-k},
\end{equation}
to be the \emph{stabilizer syndrome} corresponding to that error configuration. Here the $j$th bit $\beta_j\in\{0,1\}$, and is related to the status of $\mathcal{S}_j^c$, i.e., whether the outcome is $+1$ or $-1$, as $(-1)^{\beta_j}$. Similarly, in the case of the graphs, the graph syndrome is an $N$-bit string $\alpha\in \text{GF}(2)$, with
\begin{equation}
    \alpha = \alpha_1\alpha_2\cdots \alpha_N,
\end{equation}
where $\alpha_i\in\{0,1\}$, and are related to the status of the graph generators as $(-1)^{\alpha_i}$. Note that to each configurations of the bit-strings $\alpha$ ($\beta$), a unique decimal number between $0$ and $2^N-1$ ($2^{N-k}-1$) can be assigned.  In the rest of this paper, we use the decimal and the binary representations of the bit-strings interchangeably and the operations involving $\alpha$ and $\beta$ are assumed to be modular two.

Note that for a stabilizer code corresponding to which the equivalent graph has been identified, $\alpha$ is of length $N$, while the same for $\beta$ is $N-k$, which is  due to the addition of the $k$ logical operators $\mathcal{L}^z_i$ to the stabilizer set (Eq.~(\ref{eq:total_stabilizers})) in order to determine the equivalent graph. In QEC, the measurements of $\mathcal{L}^z_i$s are not allowed for the purpose of decoding, as they result in the collapse of the preserved logical information, prohibiting the possibility of obtaining a unique $\alpha$ from each $\beta$. This is remedied by considering the  logical syndrome $\tilde\beta$ of length $k$, given by
\begin{equation}\label{eq:log_syndrome}
\tilde\beta=\tilde\beta_1\tilde\beta_2\cdots\tilde\beta_k,
\end{equation}
with $\tilde{\beta}_j\in\{0,1\}$ where the status of $\mathcal{L}^z_j$ is  $(-1)^{\tilde\beta_j}$, and $\tilde{\beta}$ takes $2^k$ possible values. Corresponding to each configuration of logical syndrome, the total syndrome, $\gamma$, of length $N$ can be obtained by combining both stabilizer (Eqs.~(\ref{eq:stab_syndrome})) and logical (Eq.~(\ref{eq:log_syndrome})) syndromes via concatenating them in the same order as they appear in Eq.~(\ref{eq:total_stabilizers}), i.e.,  
\begin{equation}\label{eq:tot_syndrome}
\gamma=\beta\tilde\beta.
\end{equation}
Finally, using Eq.~(\ref{eq:A_S_to_A_G}) and the fact that measurement outcomes are invariant under local Clifford operations $R_l$ and $H_r$, the corresponding  $\alpha$ can be obtained from $\gamma$ as 
\begin{equation}\label{eq:tot_syndrome_to_graph_syndrome}
    \alpha=\gamma J.    
\end{equation}
In Fig.~\ref{fig:5_1_3_perfect_code} we demonstrate the equivalent graph and associated graph syndromes of two cases of the $[[5,1,3]]$ perfect code which is the smallest stabilizer code that can correct arbitrary single qubit errors. Also see Appendix.~\ref{app:5_1_3_perfect_code} for a detailed discussion. 

\subsection{Cosets and their graphical representation}
\label{subsec:cosets}

Restricting to only Pauli errors, $4^N$ possible Pauli error configurations, denoted by $\mathcal{E}$, are possible on a register of $N$ physical qubits. In the case of a graph state $\ket{G}$, each graph syndrome corresponds to an element in the graph state basis~\cite{Hein2006}, given by $\set{\ket{\alpha};\alpha=0,1,\cdots,2^N-1}$, with 
\begin{equation}
    \ket{\alpha}=Z_{\alpha} \ket{G}=\bigotimes_i (\sigma_i^z)^{\alpha_i} \ket{G},
\end{equation}
where we identify $\ket{0}=\ket{G}$. Owing to the fact that $\sigma_i^y\equiv\sigma_i^x\sigma_i^z$ up to a global phase, the action of Pauli operators on any node is equivalent to switching from one graph syndrome to another up to a global phase, as 
\begin{eqnarray}    \sigma_i^x\ket{\alpha}&=&\ket{\alpha\oplus \mathcal{N}_i}\label{eq:pauli_on_graph_x},\\
\sigma_i^y\ket{\alpha}&=&\ket{\alpha\oplus\mathcal{N}_i\oplus i}\label{eq:pauli_on_graph_y},\\    
\sigma_i^z\ket{\alpha}&=&\ket{\alpha\oplus i}\label{eq:pauli_on_graph_z}.
\end{eqnarray}
Here, in the direct sum,  $i$ $(\mathcal{N}_i)$ corresponds to the bit string with $1$ at the node (at all nodes in its neighborhood), and zeroes everywhere else (see Fig.~\ref{fig:5_1_3_perfect_code}). Therefore, an arbitrary $\mathcal{E}$ on a graph state results in one of the $2^N$ possible graph syndromes.

For an arbitrary $\mathcal{E}$, we now introduce an efficient representation in terms of ordered pair $(\mu,\nu)$ of two bit-strings, each with length $N$, as
\begin{eqnarray}
    \mu&=&\mu_1\mu_2\cdots\mu_N,\;\; 
    \nu=\nu_1\nu_2\cdots\nu_N,
\end{eqnarray}
with $\mu_i,\nu_i\in\{0,1\}$, denoting the non-trivial actions of $\sigma^x$ and $\sigma^z$ errors respectively on the individual qubits, such that
\begin{equation}
    \mathcal{E}\equiv (\mu,\nu)\to \mathcal{E}=\bigotimes_i (\sigma_i^x)^{\mu_i} (\sigma_i^z)^{\nu_i}.
\end{equation}
By virtue of Eqs.~(\ref{eq:pauli_on_graph_x})-(\ref{eq:pauli_on_graph_z}), the ordered pair  $(\mu,\nu)$ corresponding to $\mathcal{E}$, along with the graph adjacency matrix $\Gamma$, leads to\footnote{Multiplying $\mu$ with  $\Gamma$ requires treating $\mu$ to be a row vector. This is followed for all appropriate multiplications of a bit-string with a matrix.}   
\begin{eqnarray}
\mathcal{E}\ket{0}
&=&\bigotimes_i(\sigma^x_i)^{\mu_i}(\sigma^z_i)^{\nu_i} \ket{0}\nonumber\\
&=&\ket{\mu\Gamma\oplus\nu}\label{eq:gamma_mu+nu}.
\end{eqnarray}
A coset $\mathscr{E}_\alpha$ is defined as the set of all $\mathcal{E}$ resulting in the same graph syndrome $\alpha$, i.e,
\begin{equation}\label{eq:coset}
\mathscr{E}_\alpha=\set{\mathcal{E}\equiv(\mu,\nu):\mathcal{E}\ket{0}=\ket{\alpha}},   
\end{equation}
leading to
\begin{eqnarray}\label{eq:zero_alpha}
    (0,\alpha)\in \mathscr{E}_{\alpha}.    
\end{eqnarray}
Hence, starting from $(0,\alpha)$, and using Eq.~(\ref{eq:gamma_mu+nu}), the entire $\mathscr{E}_\alpha$ can be generated as
\begin{eqnarray}\label{eq:coset_k}
    \mathscr{E}_\alpha=\{(\mu,\mu\Gamma\oplus \alpha):\mu\in [0,2^N]\},
\end{eqnarray}
which has a cardinality $|\mathscr{E}_\alpha|=2^N$.

\section{Graph-aware bounded distance decoding}
\label{sec:bounded_distance}

In this Section, we discuss the methodology to decode the physical error from a syndrome for a given stabilizer code and the corresponding equivalent graph representation.  Note that the weight, $\text{w}[\mathcal{E}]\equiv\text{w}[(\mu,\nu)]$, of $\mathcal{E}$, defined by the number of qubits subjected to a non-trivial action due to $\mathcal{E}$, is given by  
\begin{equation}\label{eq:weight}
    \text{w}[\mathcal{E}]=|\{i:\mu_i\neq 0,\text{ or }\nu_i\neq 0\}|.
\end{equation}
Even without any further weight reductions, $\mathcal{E}\equiv(0,\alpha)$ already provides a correction by virtue of Eq.~(\ref{eq:zero_alpha}), conforming to the error syndrome that may or may not cause a logical error. This ensures that the graph-aware formalism never results in a decoder failure, i.e., when the decoder fails to come up with at least one correction consistent with the syndrome~\cite{Poulin2008, Panteleev2021}. The \emph{coset minimum}, denoted by 
\begin{equation}
    \mathcal{E}_{\beta,\tilde\beta,m}\equiv(\mu_m,\nu_m=\mu_m\Gamma\oplus\alpha),
\end{equation}
is the minimum weight member of $\mathscr{E}_\alpha$ (Eq.~(\ref{eq:coset})), i.e., 
\begin{eqnarray}\label{eq:coset_min}
\mathcal{E}_{\beta,\tilde\beta,m} \ni \text{w}[\mathcal{E}_{\beta,\tilde\beta,m}] =\min_{\mu} \text{w}[(\mu,\mu\Gamma\oplus\alpha)],
\end{eqnarray}
where the use of $\beta$ and $\tilde\beta$ in the subscript of $\mathcal{E}$ instead of $\alpha$ is a reminder that each $\alpha$ is a resultant of $\beta$ and $\tilde\beta$ (see Eqs.~(\ref{eq:tot_syndrome})-(\ref{eq:tot_syndrome_to_graph_syndrome})). As shown in Sec.~\ref{subsec:stab_to_graph_syndrome}, a specific stabilizer syndrome, $\beta$, leads to $2^k$ graph syndromes, $\alpha$, via different logical syndromes, $\tilde{\beta}$,  each of which possesses the coset minimum (Eq.~(\ref{eq:coset_min})). A minimization over the $2^k$ possible configurations of $\tilde\beta$ provides the \emph{global} minimum weight Pauli error, $\mathcal{E}_{\beta,m}$, corresponding to the chosen $\beta$, i.e., 
\begin{equation}\label{eq:global_min}
    \mathcal{E}_{\beta, m}\ni \text{w}[\mathcal{E}_{\beta,m}]= \min_{\tilde\beta} \text{w}[\mathcal{E}_{\beta,\tilde\beta,m}].
\end{equation}
The goal of the graph-aware decoder is to perform this minimization using $\Gamma$, while the maximum likelihood decoding (MLD) (non-degenerate decoding to be precise)~\cite{MacWilliams1977,Iyer2015,deMarti2024} is achieved by the decoder if  the set 
\begin{equation}\label{eq:MLD}
    \left\{\mathcal{E}_{\beta,m}:\text{w}[\mathcal{E}_{\beta,m}]=\min_{\tilde\beta}\left[\min_{\mu}\bigg(\text{w}[(\mu,\mu\Gamma+\alpha)]\bigg)\right], \forall \beta \right\},
\end{equation}
can be constructed by determining $\mathcal{E}_{\beta, m}$ for all possible $2^{N-k}$ possible values of $\beta$. Note that although reduced from $4^N$, the search space for the coset minimum ($2^N$) is exponential in $N$, making the problem NP-hard.

\subsection{Bounded distance decoding}
\label{subsec:bdd}

By construction, only a small fraction 
\begin{equation}
    f([[N,k,d]])=\frac{\sum_{q=0}^t{N\choose q}3^q}{2^{N-k}}\ll 1,
\end{equation}
of all possible stabilizer syndromes are corrected by an arbitrary $[[N,k,d]]$ stabilizer code with certainty, as $t$ (see Eq.~(\ref{eq:t})) is typically $\ll N$. Therefore, instead of the MLD, one can consider a bounded distance decoding (BDD)~\cite{MacWilliams1977,Liu2011,Kasai2012,Aggarwal2025,Li2010_c}, where one aim to ensure the decoding of all syndromes with the weight of the corresponding coset minima lying within a target weight $T$, i.e., 
\begin{equation}
    \text{w}[\mathcal{E}_{\beta,m}]\leq T\; \forall\text{ decoded }\beta,
\end{equation}
and any $\mathcal{E}_{\beta,m}$ with $\text{w}[\mathcal{E}_{\beta,m}]>T$ for a decoded $\beta$ is an additional benefit of the decoding process being totally unintentional. While this is, in general, a difficult task due to the inaccessibility of $\text{w}[\mathcal{E}_{\beta,m}]$ from $\beta$ alone, the equivalent graph descriptions of stabilizer codes (see Sec.~\ref{subsec:equivalent_graphs}) makes it possible. 

The Hamming weight of a bit-string, say, $\mu$, is defined by 
\begin{equation}
    \text{w}[\mu]=|\{i:\mu_i=1\}|=\sum_{i}\mu_i.
\end{equation}
with 
\begin{equation}\label{eq:coset_member_property}
    \text{w}[(\mu,\mu\Gamma\oplus \alpha)]\geq \text{w} [\mu],
\end{equation}
for every member of the coset. Therefore, starting from a specific $\alpha$, and encompassing all coset members enforcing $\text{w} [\mu]\leq T$, the coset minimum can be obtained with certainty as long as $\text{w}[\mathcal{E}_{\beta,\tilde\beta,m}]\leq T$, while 
arriving at the coset minimum using this restricted sampling strategy is not guaranteed if $\text{w}[\mathcal{E}_{\beta,\tilde\beta,m}]>T$. Therefore, modifying the sampling strategy of $\mu$, BDD can be formulated as 
\begin{equation}\label{eq:BDD_general}
    \left\{\mathcal{E}_{\beta,m}:\text{w}[\mathcal{E}_{\beta,m}]=\min_{\tilde\beta}\left[\min_{\mu,\text{w}[\mu]\leq T}\bigg(\text{w}[(\mu,\mu\Gamma+\alpha)]\bigg)\right], \forall \beta \right\},
\end{equation}
resulting in a reduced search-space cardinality, in comparison with the MLD, required to arrive at the coset minimum, as
\begin{equation}\label{eq:MLD_to_BDD_reduction}
    \sum_{q=0}^T {N\choose q}\ll 2^N. 
\end{equation}
Setting $T=N$, one retrieves the MLD from the BDD. 

It is important to note that the graph state representation and the structure of graph generators (see Eq.~(\ref{eq:graph_generators})) is crucial for the validity of Eq.~(\ref{eq:coset_member_property}), and subsequently for facilitating the BDD,  even without any prior knowledge of $\text{w}[\mathcal{E}_{\beta,\tilde\beta,m}]$. Since an $[[N,k,d]]$ stabilizer code can correct any $\mathcal{E}$ having $\text{w}[\mathcal{E}]\leq t$ with certainty, it is ensured that the set of all coset minima contains all Pauli errors with weight up to $t$\footnote{This is always true for all non-degenerate codes. However, for degenerate codes, for any error $\mathcal{E}$ with $\text{w}[\mathcal{E}]\leq t$, the associated global minima,  $\mathcal{E}_{\beta,m}$, is such that the net operation, $\mathcal{E}\mathcal{E}_{\beta,m}$, does not cause a logical error. Therefore, the set of all coset minima need not cover all Pauli errors with weight up to $t$, although it can correct all of them.}. Therefore, unless otherwise stated, we set $T=t$ for further discussions.

\subsection{Feedforward network, graph pruning and structured sampling}
\label{subsec:FFN_pruning_struct_sampling}

\begin{figure*}
    \centering
    \includegraphics[width=0.6\linewidth]{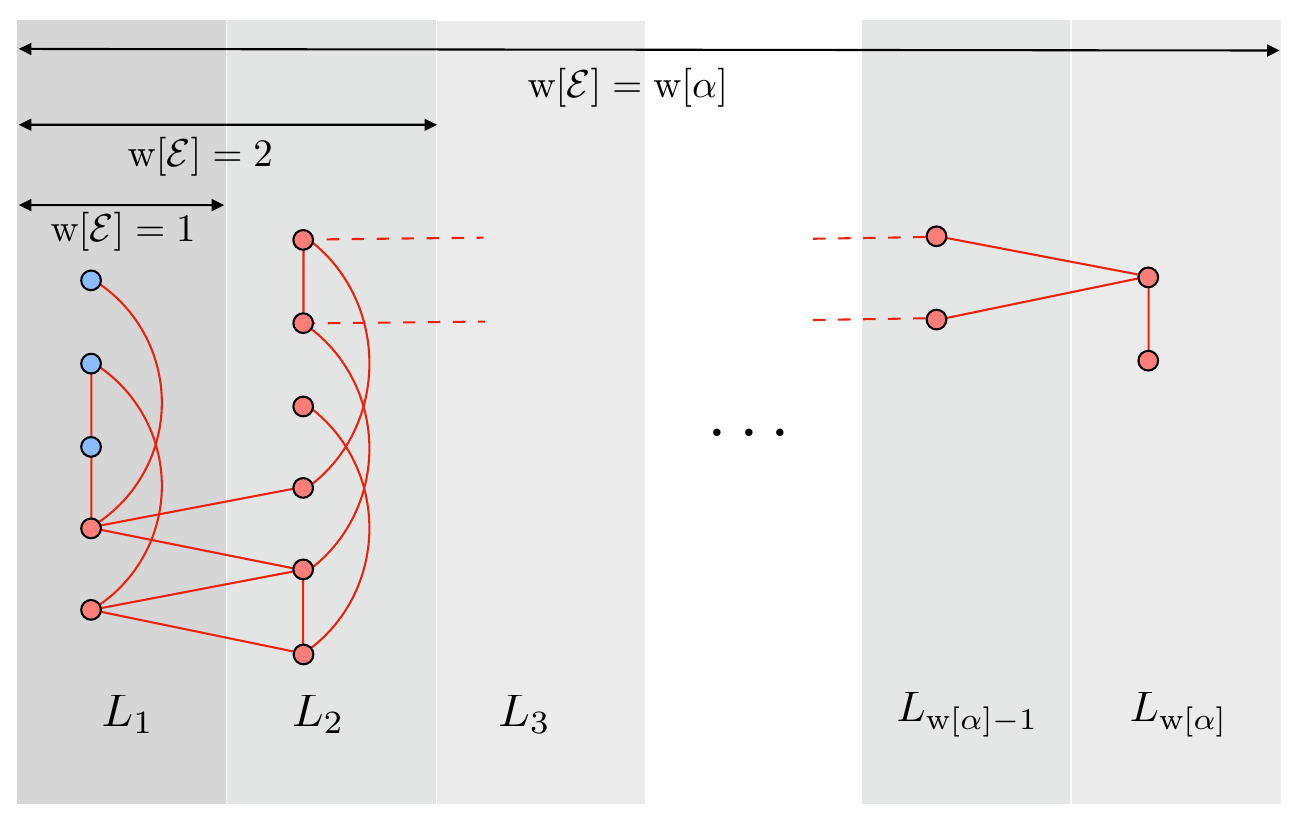}
    \caption{A demonstration of feedforward network structure of a graph. The graph syndrome nodes (shown in blue) and their neighbors belong to the left most layer $L_1$. Any later layer contains neighbors of preceding layer as well as their neighbors (See Sec.~\ref{subsec:FFN_pruning_struct_sampling}). Nodes after $L_{\text{w}[\alpha]}$ can be discarded by virtue of Eq.~(\ref{eq:coset_minimum_weight_upper_bound}).}
    \label{fig:feed_forward}
\end{figure*}

Starting from $\alpha$, the runtime in arriving at the coset minimum can be reduced further. Consider the choice of $N$ bit strings, $\overline\mu_i$s, each having $N$ bits, and unit Hamming weight with $1$ only at the $i^{th}$ position. i.e.,
\begin{equation}
    \overline\mu_i=0_10_2\cdots0_{i-1}1_i0_{i+1}\cdots 0_N.
\end{equation}
Starting from $(0,\alpha)$ (see Eqs.~(\ref{eq:coset}), (\ref{eq:zero_alpha})), one can arrive at $(\mu_m,\nu_m)$ via a linear combination of $\overline\mu_i$s as
\begin{equation}
    \mu_m=\sum_i c_i\overline\mu_i,\;\;\nu_m=\alpha\oplus\sum_i c_i\overline\mu_i\Gamma,
\end{equation}
where $c_i\in\{0,1\}$. The process of arriving at the coset minimum can be recast to either random, or structured sampling of $c_i$s with the constraint that the non-trivial combination of $c_i$s, as given by $\mu_m$, yields  the least-weight member $(\mu_m,\nu_m)$.  

We now describe certain properties of any graph that can speed up this search. Since $(0,\alpha)\in\mathscr{E}_\alpha$, for any non-trivial syndrome, the coset minimum satisfy
\begin{equation}\label{eq:coset_minimum_weight_upper_bound}
    1\leq\text{w}[(\mu_m,\nu_m)]\leq \text{w}[\alpha].
\end{equation}
Also, note that
\begin{equation}
    \text{w}[\overline\mu_i+\overline\mu_j]>\text{w}[\overline\mu_i]=\text{w}[\overline\mu_j]\;\forall\; i,j,
\end{equation}
and the weight of every coset member is lower bounded according to Eq.~(\ref{eq:coset_member_property}). Therefore, a reduction of weight from $(0,\alpha)$ is only possible via appropriate choice of $\overline\mu_i$s such that
\begin{equation}
    \text{w}[\alpha+\sum_i c_i\overline\mu_i\Gamma]<\text{w}[\alpha],
\end{equation}
thereby restricting the sampling of $c_i$s depending on the graph connectivity. 

Based on these insights, we propose the following efficient strategy. Consider a rearrangement of the original graph, $G$, with the graph syndrome, $\alpha$, into a feedforward network (FFN) (see Fig.~\ref{fig:feed_forward} for a demonstration). The first layer, $L_1$, of the FFN is constituted of the \emph{graph syndrome nodes} where $\alpha_i=1$, and their neighbors,  i.e, $L_1=\{i\}$ such that $\alpha_i=1$ or $\Gamma_{ij}=1$ with $\alpha_j=1$. Further, any successive forward layer of the FFN is constituted of nodes connected to at least one of the nodes in the preceding layer, and their neighbors, such that the $m$th layer is given by $L_{m>1}=\{i\}$ such that $\Gamma_{i,j}=1$ with $j\in L_{m-1}$ or $\Gamma_{i,j}=1$ with $k\in L_{m-1}$ and $\Gamma_{j,k}=1$. Following this, starting from $\alpha$ and using $\Gamma$, the FFN corresponding to $\alpha$ can be constructed. Due to the specific structure of the graph generators (Eq.~(\ref{eq:graph_generators})), and therefore the coset enumeration (Eq.~(\ref{eq:coset_k})), any $\mathcal{E}$ with $\text{w}[\mathcal{E}]=T$ is situated within $L_1$ and $L_{T}$ layers of the FFN. Since the weight of the coset minimum is upper-bounded by $\text{w}[\alpha]$ (Eq.~(\ref{eq:coset_minimum_weight_upper_bound})), any node appearing in the FFN beyond $L_{\text{w}[\alpha]}$ can be discarded. This is referred to as the \emph{graph pruning}, resulting in further reduction of search space in comparison with Eq.~(\ref{eq:MLD_to_BDD_reduction}).  In Fig.~\ref{fig:feed_forward}, we demonstrate an instance of a FFN structure corresponding to an arbitrary graph and graph syndrome. We point out here that the achievable reduction in runtime to arrive at the coset minimum by the graph pruning depends heavily  on the graph connectivity, and, therefore, is code-specific.

Note that the FFN structure motivates choosing a \emph{structured sampling} of $c_i$s over the random sampling, specifically since none of the $c_j \in L_{m+1}$ can assume non-trivial values for the resultant coset member to have a reduced weight unless a non-trivial value is assumed by $c_i\in L_m$. A structured sampling utilizing this possess a reduced search space in comparison with the direct sampling (see Eq.~(\ref{eq:MLD_to_BDD_reduction})). Further, the sampling of the allowed $c_i$s can be done orderly, such that the shallow layers are sampled first, followed by the deeper ones. This can be justified from the fact that under uncorrelated noise channels on the physical qubits, the low-weight errors are more likely than the higher-weight ones. 

\subsection{Performance in optimal non-CSS codes}
\label{subsec:optimal_codes}

\begin{figure}
    \centering
    \includegraphics[width=0.9\linewidth]{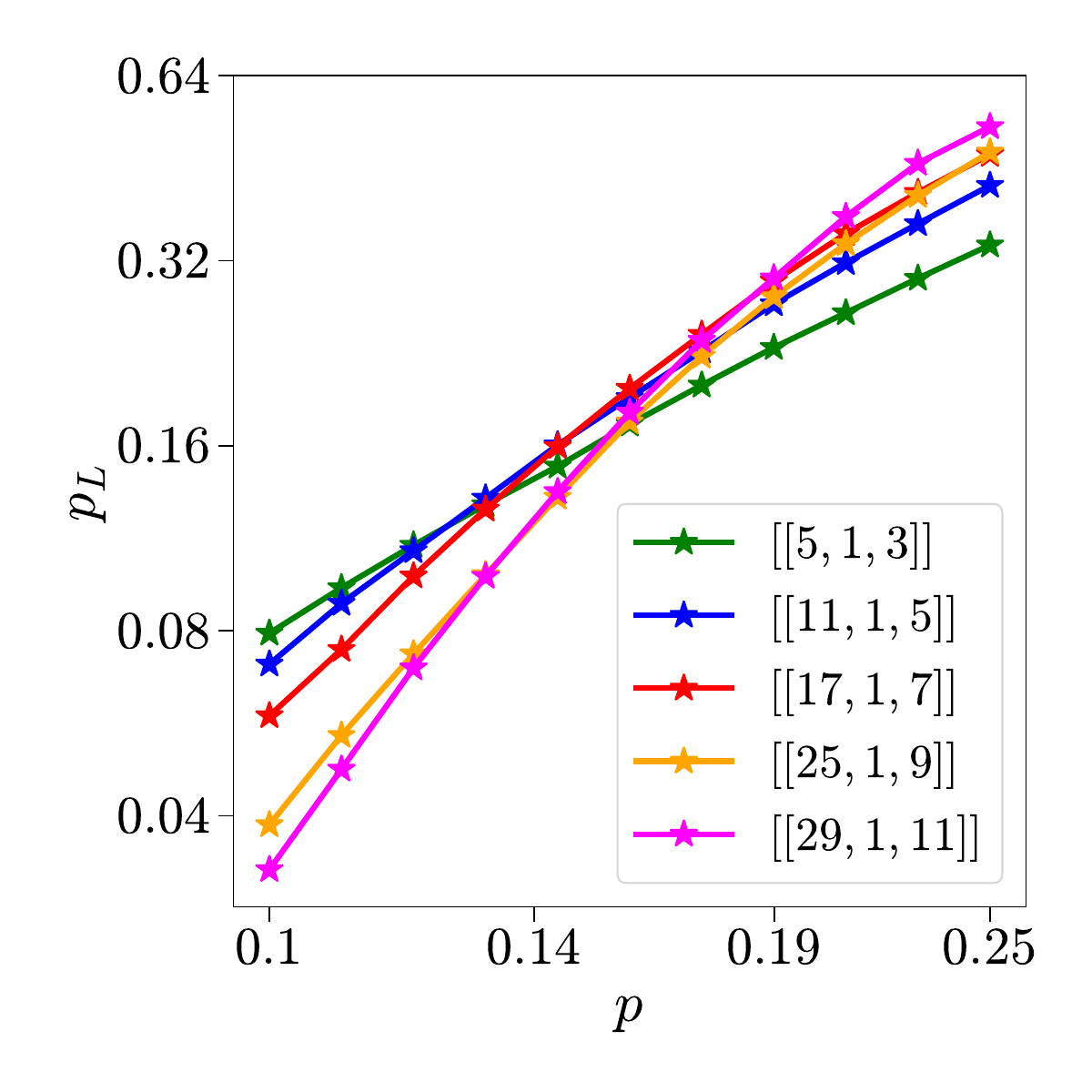}
    \caption{Log-log plot of logical error rate $p_L$ against the physical error rate $p$ in the case of optimal codes under depolarization noise corrected with BDD (see Sec.~\ref{subsec:optimal_codes} and Appendix.~\ref{app:list_of_non_CSS} for code details). Errors are sampled according to Eq.~(\ref{eq:error_model}) with $p_x=p_y=p_z=p/3$. Each of the datapoints correspond to $10^4$ occurrences of logical errors.}
    \label{fig:non_CSS}
\end{figure}

In what follows, unless otherwise stated, we use the graph pruning and structured sampling to ensure maximal efficiency of the BDD. For demonstration, we consider the single-qubit depolarization noise channels, described by
\begin{eqnarray}\label{eq:error_model}
    \Lambda_i(\rho_i)&=&(1-p_x-p_y-p_z)\rho_i+p_x\sigma^x_i\rho_i\sigma^x_i+p_y\sigma^y_i\rho_i\sigma^y_i\nonumber\\
    &&+p_z\sigma^z_i\rho_i\sigma^z_i,
\end{eqnarray}
where $\rho_i$ is the single-qubit density matrix. This is equivalent to sampling a single-qubit error from the set $\mathcal{E}\in\{I,\sigma_i^x,\sigma_i^y,\sigma_i^z\}$ with the respective probabilities $\{1-p_x-p_y-p_z,p_x,p_y,p_z\}$. For $N$ qubits, we consider the uncorrelated version of the channel, $\boldsymbol{\Lambda}=\bigotimes_i\Lambda_i(\rho)$,  acting on the $N$-qubit logical state $\rho$ belonging to the code space. We further assume that all the syndromes can be measured perfectly. 

First, we consider optimal\footnote{These codes are optimal as there exists no stabilizer codes to the date with less $N$ that achieves the same $d$~\cite{Grassl_codetables}.} non-CSS codes~\cite{Grassl_codetables} defined on $N=5,11,17,25$ and $29$ physical qubits, all encoding a single logical qubit ($k=1$) and having distances $d=3,5,7,9$, and $11$, respectively (see Appendix.~\ref{app:list_of_non_CSS} for details of the codes).  We estimate the performance of the BDD under the depolarizing channels (see Eq.~(\ref{eq:error_model})) setting $p_x=p_y=p_z=p/3$ for all qubits (see \cite{QGDecoder} for a C++ implementation of graph-aware bounded distance decoder used for the numerical simulations). In Fig.~\ref{fig:non_CSS}, we plot the variation of the logical error rate, given by $p_L=M_{L}/M$, against the physical error rate, $p$, where $M_{L}$ is the number of samples out of $M$ in which a logical error occurs. Following~\cite{Bravyi2014}, we fix $M_{L}=10^4$, and evaluate $M$ for a given value of $p$ to obtain $p_L$.  
 
It is worthwhile to point out that for a given physical error-rate $p$, the weight of the errors present is $pN$ on average, while the BDD strategy ensures correction of errors up to a weight $t$. Therefore, $p=t/N$ provides a boundary between the range $0\leq p\leq t/N$ and $p>t/N$ of $p$, where error correction is more and less probable respectively. Specifically, the values of $t/N$ for these codes are $0.2,0.181,0.176,0.16$ and $0.172$, respectively, for $d=3,5,7,9$ and $11$. Note that with increasing $d$, $p_L$ diminishes faster below near a specific value of $p$ (approximately $p=0.15$, see Fig.~\ref{fig:non_CSS}) in agreement with the boundary $p=t/N$. By quantum singleton bound~\cite{Knill1997,Calderbank1997,Gottesman2009} for codes with $k=1$, 
\begin{equation}
    \frac{t}{N}\leq \frac{1}{4}-\frac{1}{4N},
\end{equation}
which, in the limit $N\rightarrow\infty$, leads to $t/N= 0.25$ as the maximum attainable limit, which is consistent with the observations. We stress here that throughout the decoding, no instances of logical errors with $\text{w}[\mathcal{E}]\leq t$ occur.

\section{bounded distance decoding of CSS codes}
\label{sec:CSS}

\begin{figure*}
    \centering
    \includegraphics[width=0.7\linewidth]{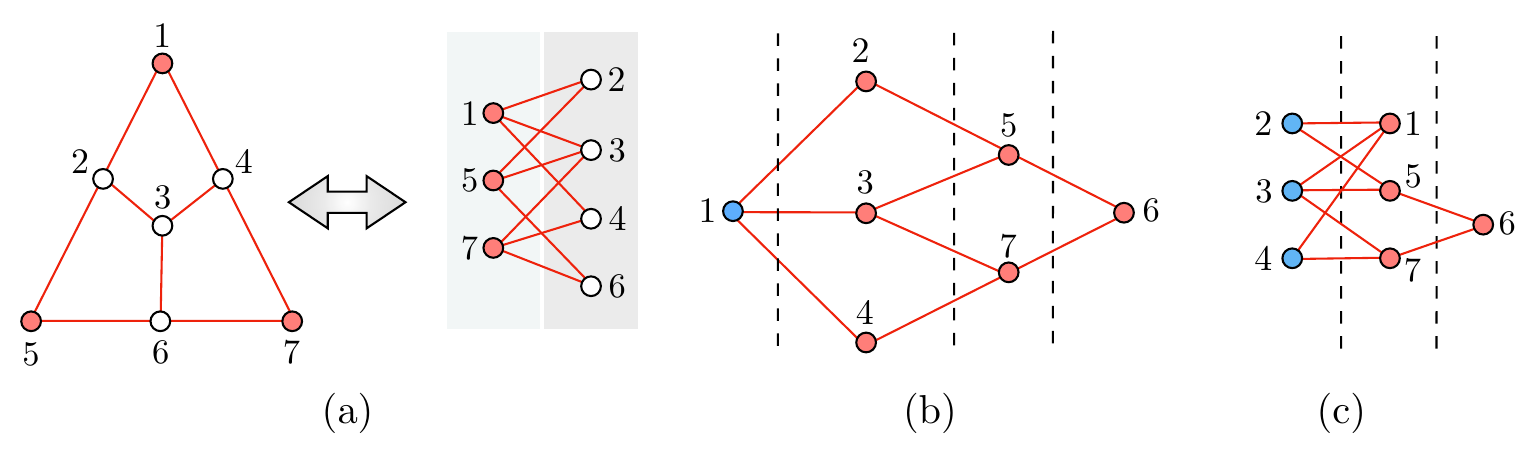}
    \caption{(a) The equivalent bipartite graph associated with smallest $[[7,1,3]]$ CSS code. The white nodes indicates Hadamard operations. See Appendix.~\ref{app:7_1_3_color_code} for the stabilizers and a detailed discussion. A demonstration of graph syndrome and the corresponding feedforward network corresponding to the trivial logical syndrome in the case of (c) $\sigma^z_1$ (d) $\sigma^x_1$ error.}
    \label{fig:7_1_3_CSS}
\end{figure*}

In CSS codes, the stabilizer operators $\mathcal{S}^c$ are either $\sigma^x$-type, denoted by $\mathcal{S}^{c,x}$, or $\sigma^z$-type, given by  $\mathcal{S}^{c,z}$, leading to 
\begin{equation}\label{eq:CSS_check_matrix_property}
    \mathcal{Z}_{1l}=0,\hspace{0.3cm}\mathcal{Z}_{1r}=0,
\end{equation}
in terms of the check matrix (Eq.~(\ref{eq:A_Sp_l_r})). Since $\mathcal{L}^z_j$ $(\mathcal{L}^x_j)$ anticommutes with $\mathcal{L}^x_j$ $(\mathcal{L}^z_j)$, but commute with all $\mathcal{S}^c$s, $\mathcal{L}^z_j$ $(\mathcal{L}^x_j)$ having the minimum weight is constituted of only $\sigma^z$ $(\sigma^x)$ operators.  Unless otherwise stated, in all subsequent discussions, we only consider the logical operators, $\mathcal{L}^z_j$ and $\mathcal{L}^x_j$, having the minimum weight. The classification of $\mathcal{S}^c$s into $\sigma^x$- or  $\sigma^z$-type  warrants independent treatments of the corrections of these types of errors. Let us consider a  $\sigma^x$ error configuration, $\mathcal{E}^x$, with weight $\text{w}[\mathcal{E}^x]\leq t$, and assume that the coset minimum, $\mathcal{E}_{\beta,0,m}$ (see Eq.~(\ref{eq:coset_min})), with the trivial logical syndrome $\tilde\beta=0$, has weight
\begin{equation}
    \text{w}[\mathcal{E}_{\beta,0,m}]\leq t=\frac{d-1}{2}.
\end{equation}
For any non-trivial logical syndrome $\tilde\beta\neq0$, due to the specific structure of $\mathcal{L}_j^x$ and using the fact that $\text{w}[\mathcal{L}_j^x]\geq d\forall j$ for codes with distance $d$ codes, 
\begin{eqnarray}\label{eq:CSS_weight_condition}
    \text{w}[\mathcal{E}_{\beta,\tilde\beta\neq0,m}]&=&\text{w}\left[\prod_j(\mathcal{L}_j^x)^{\tilde\beta_j}\mathcal{E}_{\beta,0,m}\right]\nonumber\\
    &\geq& \frac{d+1}{2}\nonumber \\
    &>&\text{w}[\mathcal{E}_{\beta,0,m}].
\end{eqnarray}
Therefore, for all $\mathcal{E}^x$ with $\text{w}[\mathcal{E}^x]\leq t$, the global minimum  in Eq.~(\ref{eq:global_min}) occurs for the graph syndrome corresponding to the trivial logical syndrome, $\mathcal{E}_{\beta,0,m}$. Hence, the minimization in Eq.~(\ref{eq:global_min}) can be discarded altogether. Note that with similar considerations for $\mathcal{L}_j^z$s,  errors of $\sigma^z$-type can be shown to have the same property, and the $\sigma^y$-errors can be treated as a combination of $\sigma^x$ and $\sigma^z$ errors with a similar result. In summary, in the case of the BDD of CSS codes with a target weight $t$,
\begin{equation}\label{eq:syndrome_property_CSS}
    \mathcal{E}_{\beta,m}= \mathcal{E}_{\beta,0,m} \forall \beta.
\end{equation} 

Further, Eq.~(\ref{eq:CSS_check_matrix_property}) results in (see Eqs.~(\ref{eq:C}) and (\ref{eq:Uz_arbitrary}))
\begin{equation}\label{eq:CSS_R_l}
    R_l=I,
\end{equation}
indicating that the local Clifford $H_r$ alone connects the graph state $\ket{G}$ to the CSS stabilizer state $\ket{\mathcal{S}}$. Also leads to a simplified form of the adjacency matrix (see Eqs.~(\ref{eq:C}) and (\ref{eq:arbitrary_stabilizer_adj})), given by 
\begin{equation}\label{eq:CSS_bpartite_adj}
    \Gamma=\begin{pmatrix}
        0 & B^T\\
        B & 0\\
    \end{pmatrix},
\end{equation}
where $B=\mathcal{X}_{1r}\mathcal{X}_{1l}^{-1}$. This corresponds to a bipartite graph with links existing only between left and right nodes, and no links within the left and the right nodes themselves. Let $N_l$ $(N_r)$ be the number of nodes in the left (right) partition of the graph. In the case of the graph syndrome associated with the trivial logical syndrome, post application of $H_r$, the $\sigma^z$-errors ($\sigma^x$-errors) on the left nodes remain unchanged while the same on the right nodes transform to $\sigma^x$-errors ($\sigma^z$-errors). Since the graph is bipartite (see Eqs.~(\ref{eq:CSS_R_l})-(\ref{eq:CSS_bpartite_adj})), every $\sigma^z$-error ($\sigma^x$-error) results in a graph syndrome with non-trivial nodes only among the $N_l$ left nodes ($N_r$ right nodes). In Fig.~\ref{fig:7_1_3_CSS}, we demonstrate this property of the graph syndrome in the case of the smallest instance of the CSS code, namely, the $[[7,1,3]]$ code (see Appendix.~\ref{app:7_1_3_color_code} for details). 

Since each $\mathcal{E}^x$ error results in the graph syndrome $\alpha_r$ on $N_r$ nodes (see the above discussion), given Eq.~(\ref{eq:syndrome_property_CSS}), BDD (Eq.~(\ref{eq:BDD_general})) for the CSS codes reduces to
\begin{eqnarray}\label{eq:BDD_CSS}
    \left\{\mathcal{E}_{\beta,m} : \text{w}[\mathcal{E}_{\beta,m}]= \min_{\mu_r,\text{w}[\mu_r]\leq t} \text{w}[(\mu_l,\alpha_r\oplus\mu_l\Gamma^{r\to l})]\right\}.
\end{eqnarray}
Here, the subscripts $r$ and $l$  associated with the bit-strings denote right and left nodes respectively,  indicating the partition of $G$ to which the corresponding nodes belong. On the other hand, $\Gamma^{r\to l}=B^T$ refers to the off-diagonal block of $\Gamma$, representing how the right nodes are incident on to the left nodes. In comparison to BDD for non-CSS codes, this already results in a reduction of search space in the case of a CSS code of same size, as
\begin{equation}
    \sum_{q=0}^t {N_r\choose q}\ll \sum_{q=0}^t {N\choose q}.
\end{equation}
Further, when the graph syndrome is arranged in the FFN structure, only even layers ($L_2,L_4,\cdots$) contribute to the non-trivial $c_i$s (see Fig.~\ref{fig:7_1_3_CSS}(c)-(d)) facilitating reduction of the weight of the coset member, since the graph syndrome belongs to the right nodes. Note that the graph pruning and structured sampling strategies (see  Sec.~\ref{subsec:FFN_pruning_struct_sampling}) can also be applied for the CSS codes, resulting in further search space reductions. 

\subsection{Decoding the color and the surface codes}
\label{subsec:color+surface}

\begin{figure*}
    \centering
    \includegraphics[width=0.6\linewidth]{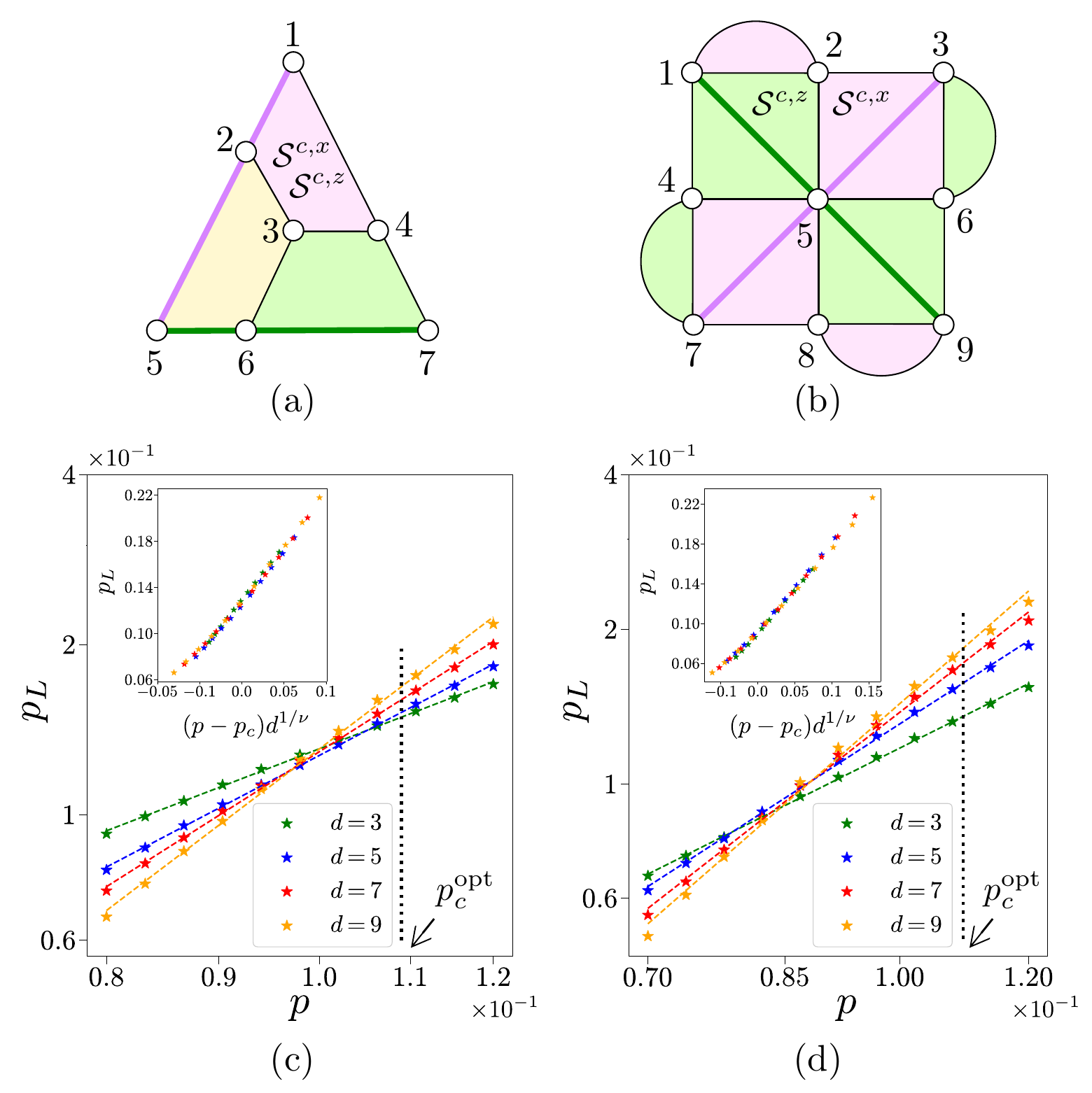}
    \caption{Schematics of $d=3,k=1$ (a) triangular color and (b) rotated surface codes with $N=7$ and $N=9$ physical qubits respectively. The thick pick and green lines represent the logical $z$ and $x$ operators respectively. The log-log plot of $p_L$ against $p$ for code distances three to nine of (c) triangular color (d) rotated surface codes. Both of them are under bit flip channel corresponding to $p_y=p_z=0$ of Eq.~(\ref{eq:error_model}). The vertical black dotted line indicates the theoretical known optimal threshold noise strengths $p_c^{opt}=0.109(2)$\cite{Katzgraber2009} ($p_c^{opt}=0.1094(2)$\cite{Dennis2002}) of color (surface) code. The inset shows the data collapse (Eq.~(\ref{eq:data_collapse})) obtained near the BDD threshold physical error probability $p_c$ for (c) color and (d) surface code by optimizing the values of $p_c$ and $\nu$. The obtained values are $p_c=0.098$ ($p_c=0.084$) and $\nu=1.503$ ($\nu=1.500$) for color (surface) code.}
    \label{fig:color_surface}
\end{figure*}

As a demonstration, we perform the BDD for the triangular color code~\cite{Bombin2006,Landahl2011,Katzgraber2009,Thomsen2024,Lee2025,Berent2024} and the rotated surface code~\cite{Kitaev2003,Fowler2012,Horsman2012,Litinski2019,Acharya2025}. These are respectively $[[(3d^2+1)/4,1,d]]$ and $[[d^2,1,d]]$ code families, and are the leading candidates for fault-tolerant quantum technologies~\cite{Campbell2017,Fowler2012,Terhal2015,Litinski2019,Thomsen2024}. The triangular color (rotated surface) code has three (two) types of plaquettes, designated by three (two) different colors, namely, pink, yellow and green (pink and green), as shown in Fig.~\ref{fig:color_surface}(a) (Fig.~\ref{fig:color_surface}(b)). In the color code, each plaquette corresponds to both $\mathcal{S}^{c,x}$ and $\mathcal{S}^{c,z}$ stabilizers acting non-trivially on the qubits belonging to the plaquette, while in the surface code, pink (green) plaquettes correspond only to $\mathcal{S}^{c,x}$ ($\mathcal{S}^{c,z}$). Note that $d=3$ triangular color code, i.e, the $[[7,1,3]]$ code is the smallest CSS code, which is already discussed above (see Fig.~\ref{fig:7_1_3_CSS} and Appendix~\ref{app:7_1_3_color_code}).  

We perform the BDD of both the color and the surface codes under the bit-flip channel, which corresponds to the special case $p_y=p_z=0$ of the noise channel given in Eq.~(\ref{eq:error_model}) (see \cite{QGDecoder} for a C++ implementation of graph-aware bounded distance decoder used for the numerical simulations). The BDD ensures that all errors with weight up to $t$ are corrected with certainty. While this results in a search space reduction as established in Sec.~\ref{sec:bounded_distance}, the search space dimension quickly goes beyond the scope of our available numerical resource with increasing $d$, restricting our numerical simulations within codes with $d\leq 9$.  For estimating $p_c$ using BDD, we explore the data collapse near the crossover regime according to the reported finite size scaling anzatz~\cite{English2025,English2025_2}, given by
\begin{equation}\label{eq:data_collapse}
    p_L=f[(p-p_c)d^{1/\nu}].
\end{equation}
We carry out simultaneous optimization over both $\nu$ and $p_c$, minimizing the magnitude of the error in least square fit of the data to a polynomial finite-size scaling function $f$~\cite{Sandvik2010}. For the color (surface) code, the threshold probability, $p_{c}^{\text{opt}}$, achievable by optimal decoding is known to be $p_c^{\text{opt}}=0.109(2)$~\cite{Katzgraber2009} ($p_c^{\text{opt}}=0.1094(2)$~\cite{Dennis2002}), while the BDD offers $p_c=0.098$ ($p_c=0.084$), as shown in  Fig.~\ref{fig:color_surface}(c)-(d), which are comparable to $p_c^{\text{opt}}$.

In the insets of Fig.~\ref{fig:color_surface}(c)-(d), we plot $p_L$ against the rescaled physical error rate in the vicinity of $p_c$, demonstrating a data collapse. We obtain the value of the exponent in Eq.~(\ref{eq:data_collapse}) to be $\nu=1.503$ ($\nu=1.500$) for the color (surface) code. Note that the color (surface) code under the bit-flip channel on all qubits maps to the two-dimensional (2D) random three-body Ising model (RTBIM)~\cite{Katzgraber2009} (random-bond Ising model (RBIM)~\cite{Dennis2002}). In the case of the surface code, the obtained values of $\nu$ are comparable to the values ($\nu=1.33(3)$~\cite{Honecker2001}, $\nu=1.5(1)$~\cite{Merz2002}) reported for the 2D RBIM in literature~\cite{Honecker2001,Merz2002}. However, for the color code, the numerical investigation of the 2D RTBIM suggests $\nu=2/3$~\cite{Katzgraber2009}. The lack of agreement in the exponent for the color code and the 2D RTBIM can potentially be caused by the limited system sizes under investigation.

\section{Conclusion and outlook}
\label{sec:conclusion}

In this paper, we have developed a framework for bounded distance decoding of arbitrary stabilizer codes exploiting the idea that an arbitrary stabilizer state can be transformed to a graph state via local Clifford operations. While the methodology to obtain the representative graphs for arbitrary stabilizer states exists, we give a prescription for arriving at the equivalent graph syndromes from the stabilizer syndromes. We propose the bounded distance decoding as a generic decoding strategy for all stabilizer codes, from which the maximum likelihood decoding emerges  as a limiting case. The designed strategy enables a reduction in search space en route to arrive at the minimum weight error configuration corresponding to a syndrome, while ensuring that errors up to the target weight are always corrected. Additional insights from the underlying graph structure facilitates further search space reductions via graph pruning and structured sampling. We also provide an open-source numerical package \href{https://github.com/harikrishnan9779/QGDecoder}{QGDecoder} for the graph-aware bounded distance decoding.

We demonstrate the performance of the decoder on optimal non-CSS codes up to distance eleven. Also, for codes belonging to the  CSS family, our strategy extracts further advantages in search-space reduction exploiting the bipartite properties of the underlying graphs corresponding to the CSS code. Application of the decoder to the color and surface CSS code families reveals  near optimal performance, establishing the decoder on a stronger footing. These initial performance parameters of the designed decoder calls for follow-up investigations into improving the design via developing further runtime reduction strategies.  Also, realistic hardware-aware extensions of the idea, for example, towards handling imperfect syndrome measurements~\cite{Dennis2002,Fowler2012} and presence of circuit level noise~\cite{Fowler2012,Terhal2015,Lee2025} remains to be explored.

\acknowledgements 
H.K.J. thanks the Prime Minister's Research Fellowship (PMRF) program, Government of India, for the financial support. A.K.P acknowledges the support from the Anusandhan National Research Foundation (ANRF) of the Department of Science and Technology (DST), India, through the Core Research Grant (CRG) (File No. CRG/2023/001217, Sanction Date 16 May 2024).

\onecolumngrid 

\appendix

\section{Extracting equivalent graphs of stabilizer codes}
\label{app:stab_to_graph}

Here we provide necessary details for obtaining the representative graphs from a stabilizer code. The union of the set of $N-k$ linearly independent stabilizer operators, $\mathcal{S}^c$ (Eq.~(\ref{eq:stabilizers})), and the set of $k$ logical operators, $\mathcal{L}^z$ (Eq.~(\ref{eq:logicals})), is given by 
\begin{equation}
    \mathcal{S}=\mathcal{S}^c\cup \mathcal{L}^z,
\end{equation}
defining a unique stabilizer state, $\ket{\mathcal{S}}$, among the codewords of the stabilizer code. Since the code space of a stabilizer code contains $2^k$ states, we consider the common $+1$ eigenstate of all of the $k$ logical operators. Note that this choice is completely arbitrary, and any other choice differs from this state via local Clifford operations only. The check matrix of any $N$-qubit stabilizer state, $\ket{\mathcal{S}}$, has the form
\begin{equation}
    A_\mathcal{S}=\begin{pmatrix}
        \mathcal{Z}\\\mathcal{X}
    \end{pmatrix},
\end{equation}
with $\mathcal{Z}$ and $\mathcal{X}$ being $N\times N$ matrices,  given by,
\begin{eqnarray}
    \mathcal{Z}_{ij}&=&0,\mathcal{X}_{ij}=0 \hspace{0.2cm}\text{iff}\hspace{0.2cm}I_i\in\mathcal{S}_j,\;\;\;\;
    \mathcal{Z}_{ij}=1,\mathcal{X}_{ij}=0 \hspace{0.2cm}\text{iff}\hspace{0.2cm}\sigma_i^z\in\mathcal{S}_j,\nonumber\\
    \mathcal{Z}_{ij}&=&0,\mathcal{X}_{ij}=1 \hspace{0.2cm}\text{iff}\hspace{0.2cm}\sigma_i^x\in\mathcal{S}_j,\;\;\;\;\mathcal{Z}_{ij}=1,\mathcal{X}_{ij}=1 \hspace{0.2cm}\text{iff}\hspace{0.2cm}\sigma_i^y\in\mathcal{S}_j.
\end{eqnarray}

A graph state, $\ket{G}$, associated with a simply-connected undirected graph, $G$, defined on $N$ nodes, $V=\{1,2,\cdots,N\}$, connected by the set of links, $L=\{(i,j):\forall (i,j)\in G\}$, is also a stabilizer state, which is the common $+1$ eigenstate of the stabilizers  referred to as the graph generators. Each graph $G$ with $N$ nodes corresponds to $N$ graph generators, each of which is associated to a node in the graph (see Eq.~(\ref{eq:graph_generators})). Using the specific form of the graph generators,  the check matrix for an arbitrary graph state takes the form
\begin{equation}
    A_g=\begin{pmatrix}
        \Gamma\\I
    \end{pmatrix},
\end{equation}
where $\Gamma$ is the adjacency matrix of the corresponding graph, $G$, given by Eq.~(\ref{eq:adjacency_matrix}). Since all stabilizer states are local Clifford equivalent to a graph state~\cite{Amaro2020,VandenNest2004}, it is possible to find a transformation $A_\mathcal{S}\to A_g$ for arbitrary $A_\mathcal{S}$. Since the stabilizer generators are all linearly independent, $A_\mathcal{S}$ is of full rank $N$. Assuming rank of $\mathcal{X}$ to be $n$, starting from Eq.~(\ref{eq:A_S_arbitrary}), it is always possible, by rearrangements and Gaussian elimination~\cite{MacWilliams1977} among the columns, to arrive at the following form of the check matrix:
\begin{equation}\label{eq:A_Sp_arbitrary}
    A_\mathcal{S}^\prime=\begin{pmatrix}
        \mathcal{Z}_1 & \mathcal{Z}_2\\
        \mathcal{X}_1 & 0
    \end{pmatrix},
\end{equation}
Here, $\mathcal{X}_1$ is a full rank $N\times n$ matrix, and  $\mathcal{Z}_1$ and $\mathcal{Z}_2$ are of sizes $N\times n$ and $N\times (N-n)$,  respectively. Since $\mathcal{X}$ is rank $n$, it is always possible to choose an $n\times n$ invertible sub matrix $\mathcal{X}_{1l}$, while the remaining $(N-n)\times n$ sub matrix is referred to as $\mathcal{X}_{1r}$. We call the qubits belonging to $\mathcal{X}_{1l}$ $(\mathcal{X}_{1r})$ to be the left (right) nodes. Maintaining the left-right qubit-ordering, $\mathcal{Z}$ also naturally divides into sub matrices as 
\begin{equation}
        A_\mathcal{S}^\prime=\begin{pmatrix}
        \mathcal{Z}_{1l} & \mathcal{Z}_{2l}\\
        \mathcal{Z}_{1r} & \mathcal{Z}_{2r}\\
        \mathcal{X}_{1l} & 0\\
        \mathcal{X}_{1r} & 0
    \end{pmatrix}.
\end{equation}
The graph corresponding to $A_\mathcal{S}^\prime$ can be extracted as~\cite{Amaro2020}
\begin{equation}
    A_g=R_lH_rA_\mathcal{S}^\prime J,
\end{equation}
where $H_r$ corresponds to Hadamard operations on all of the right nodes, given by 
\begin{equation}
    H_r=\begin{pmatrix}
        I & 0 & 0 & 0\\
        0 & 0 & 0 & I\\
        0 & 0 & I & 0\\
        0 & I & 0 & 0\\
    \end{pmatrix},
\end{equation}
and $R_l$ is the single-qubit phase gate on the selected left nodes, given by 
\begin{equation}\label{eq:Uz_arbitrary}
    R_l=\begin{pmatrix}
        I & 0 & \text{diag}[C] & 0\\
        0 & I & 0 & 0\\
        0 & 0 & I & 0\\
        0 & 0 & 0 & I
    \end{pmatrix},
\end{equation}
for which $C_{ii}=1$. Here, we define
\begin{eqnarray}
    \label{eq:B}
    B&=&\mathcal{X}_{1r}\mathcal{X}_{1l}^{-1},\nonumber\\
    C&=&[\mathcal{Z}_{1l}+(\mathcal{X}_{1l}^T)^{-1}\mathcal{X}_{1r}^T\mathcal{Z}_{1r}]\mathcal{X}_{1l}^{-1},\label{eq:C}
\end{eqnarray}
in terms of which the adjacency matrix of the graph associated with the stabilizer state is given by~\cite{Amaro2020}
\begin{equation}\label{eq:arbitrary_stabilizer_adj}
    \Gamma=\begin{pmatrix}
        C+\text{diag}[C] & B^T\\
        B & 0\\
    \end{pmatrix}.
\end{equation}
Note that the matrix
\begin{equation}\label{eq:arbitrary_stabilizer_J}
    J=\begin{pmatrix}
        \mathcal{X}_{1l}^{-1} & 0\\
        \mathcal{Z}_{2r}^{-1}\mathcal{Z}_{1r}\mathcal{X}_{1l}^{-1} & \mathcal{Z}_{2r}^{-1}
    \end{pmatrix}
\end{equation}
provides the appropriate recombination of the stabilizers (Eq.~(\ref{eq:A_Sp_arbitrary})), such that the succeeding layer of local Cliffords, $R_l H_r$, results in the graph generators $A_g$.

\begin{table*}[]
\begin{tabular}{|p{0.3\textwidth}|p{0.4\textwidth}|}
\hline
\hspace{0.08\textwidth}$[[11,1,5]]$ \textbf{Code} & \hspace{0.15\textwidth}$[[17,1,7]]$ \textbf{Code} \\[1ex]
\begin{tabular}[c]{@{}l@{}}
\hspace{1cm}$\mathcal{S}_1^c = \sigma_1^x \sigma_6^x \sigma_7^z \sigma_8^z \sigma_{10}^x \sigma_{11}^x$\\
\hspace{1cm}$\mathcal{S}_2^c = \sigma_1^z \sigma_6^z \sigma_7^x \sigma_8^y \sigma_9^y \sigma_{11}^x$\\
\hspace{1cm}$\mathcal{S}_3^c = \sigma_2^x \sigma_6^x \sigma_7^z \sigma_8^y \sigma_9^z \sigma_{11}^y$\\
\hspace{1cm}$\mathcal{S}_4^c = \sigma_2^z \sigma_6^z \sigma_7^z \sigma_8^z \sigma_9^y \sigma_{10}^y$\\
\hspace{1cm}$\mathcal{S}_5^c = \sigma_3^x \sigma_6^x \sigma_7^z \sigma_8^x \sigma_9^x \sigma_{10}^z$\\
\hspace{1cm}$\mathcal{S}_6^c = \sigma_3^z \sigma_6^z \sigma_7^y \sigma_9^y \sigma_{10}^z \sigma_{11}^z$\\
\hspace{1cm}$\mathcal{S}_7^c = \sigma_4^x \sigma_6^x \sigma_7^y \sigma_9^x \sigma_{10}^x \sigma_{11}^y$\\
\hspace{1cm}$\mathcal{S}_8^c = \sigma_4^z \sigma_6^z \sigma_8^z \sigma_9^x \sigma_{10}^z \sigma_{11}^x$\\
\hspace{1cm}$\mathcal{S}_9^c = \sigma_5^x \sigma_6^x \sigma_7^x \sigma_9^z \sigma_{10}^z \sigma_{11}^x$\\
\hspace{1cm}$\mathcal{S}_{10}^c = \sigma_5^z \sigma_6^z \sigma_8^y \sigma_9^z \sigma_{10}^y \sigma_{11}^z$\\
\hspace{1cm}$\mathcal{L}^z=\sigma_2^z\sigma_5^x\sigma_6^y\sigma_7^z\sigma_8^y$\\
\hspace{1cm}$\mathcal{L}^x=\sigma_2^x\sigma_3^x\sigma_7^z\sigma_9^x\sigma_{11}^x$
\end{tabular}
&
\begin{tabular}[c]{@{}l@{}}
\vspace{1pt}\\
\hspace{1cm}$\mathcal{S}_1^c = \sigma_1^x \sigma_9^x \sigma_{10}^z \sigma_{11}^y \sigma_{12}^y \sigma_{15}^y \sigma_{16}^y \sigma_{17}^z$\\ 
\hspace{1cm}$\mathcal{S}_2^c = \sigma_1^z \sigma_9^z \sigma_{10}^y \sigma_{11}^x \sigma_{12}^x \sigma_{15}^x \sigma_{16}^x \sigma_{17}^y$\\
\hspace{1cm}$\mathcal{S}_3^c = \sigma_2^x \sigma_9^z \sigma_{10}^z \sigma_{11}^y \sigma_{12}^z \sigma_{13}^y \sigma_{15}^x \sigma_{16}^z$\\
\hspace{1cm}$\mathcal{S}_4^c = \sigma_2^z \sigma_9^y \sigma_{10}^y \sigma_{11}^x \sigma_{12}^y \sigma_{13}^x \sigma_{15}^z \sigma_{16}^y$\\
\hspace{1cm}$\mathcal{S}_5^c = \sigma_3^x \sigma_{10}^z \sigma_{11}^z \sigma_{12}^y \sigma_{13}^z \sigma_{14}^y \sigma_{16}^x \sigma_{17}^z$\\
\hspace{1cm}$\mathcal{S}_6^c = \sigma_3^z \sigma_{10}^y \sigma_{11}^y \sigma_{12}^x \sigma_{13}^y \sigma_{14}^x \sigma_{16}^z \sigma_{17}^y$\\
\hspace{1cm}$\mathcal{S}_7^c = \sigma_4^x \sigma_9^z \sigma_{10}^y \sigma_{11}^y \sigma_{12}^y \sigma_{13}^y \sigma_{14}^z \sigma_{15}^z \sigma_{16}^x \sigma_{17}^z$\\
\hspace{1cm}$\mathcal{S}_8^c = \sigma_4^z \sigma_9^y \sigma_{10}^x \sigma_{11}^x \sigma_{12}^x \sigma_{13}^x \sigma_{14}^y \sigma_{15}^y \sigma_{16}^z \sigma_{17}^y$\\
\hspace{1cm}$\mathcal{S}_9^c = \sigma_5^x \sigma_9^z \sigma_{10}^x \sigma_{11}^z \sigma_{12}^z \sigma_{13}^y \sigma_{14}^y \sigma_{15}^y \sigma_{16}^y \sigma_{17}^z$\\
\hspace{1cm}$\mathcal{S}_{10}^c = \sigma_5^z \sigma_9^y \sigma_{10}^z \sigma_{11}^y \sigma_{12}^y \sigma_{13}^x \sigma_{14}^x \sigma_{15}^x \sigma_{16}^x \sigma_{17}^y$\\
\hspace{1cm}$\mathcal{S}_{11}^c = \sigma_6^x \sigma_9^z \sigma_{10}^x \sigma_{12}^y \sigma_{13}^z \sigma_{14}^y \sigma_{15}^z \sigma_{16}^z$\\
\hspace{1cm}$\mathcal{S}_{12}^c = \sigma_6^z \sigma_9^y \sigma_{10}^z \sigma_{12}^x \sigma_{13}^y \sigma_{14}^x \sigma_{15}^y \sigma_{16}^y$\\
\hspace{1cm}$\mathcal{S}_{13}^c = \sigma_7^x \sigma_{10}^z \sigma_{11}^x \sigma_{13}^y \sigma_{14}^z \sigma_{15}^y \sigma_{16}^z \sigma_{17}^z$\\
\hspace{1cm}$\mathcal{S}_{14}^c = \sigma_7^z \sigma_{10}^y \sigma_{11}^z \sigma_{13}^x \sigma_{14}^y \sigma_{15}^x \sigma_{16}^y \sigma_{17}^y$\\
\hspace{1cm}$\mathcal{S}_{15}^c = \sigma_8^x \sigma_9^z \sigma_{10}^y \sigma_{11}^y \sigma_{14}^y \sigma_{15}^y \sigma_{16}^z \sigma_{17}^x$\\
\hspace{1cm}$\mathcal{S}_{16}^c = \sigma_8^z \sigma_9^y \sigma_{10}^x \sigma_{11}^x \sigma_{14}^x \sigma_{15}^x \sigma_{16}^y \sigma_{17}^z$\\
\hspace{1cm}$\mathcal{L}^z = \sigma_2^x \sigma_5^z \sigma_6^y \sigma_7^y \sigma_8^x \sigma_{12}^z \sigma_{17}^x$\\
\hspace{1cm}$\mathcal{L}^x = \sigma_1^y \sigma_3^y \sigma_8^x \sigma_{12}^y \sigma_{13}^z \sigma_{14}^z \sigma_{15}^x$\\
\vspace{1pt}
\end{tabular}
\\ \hline
\end{tabular}

\caption{Stabilizers for the $[[11,1,5]]$ and $[[17,1,7]]$ codes from the non-CSS family.}\label{tab:non_CSS_1}
\end{table*}

\section{The \texorpdfstring{$[[5,1,3]]$}{[[5,1,3]]} perfect code}
\label{app:5_1_3_perfect_code}

The stabilizers of the $[[5,1,3]]$ perfect code are
\begin{eqnarray}
\mathcal{S}_1^c&=&\sigma_1^x\sigma_2^z\sigma_3^z\sigma_4^x,\;\;
\mathcal{S}_2^c=\sigma_2^x\sigma_3^z\sigma_4^z\sigma_5^x,\;\;
\mathcal{S}_3^c=\sigma_1^x\sigma_3^x\sigma_4^z\sigma_5^z,\;\;
\mathcal{S}_4^c=\sigma_1^z\sigma_2^x\sigma_4^x\sigma_5^z,    
\end{eqnarray}
and the logical operators are given by 
\begin{eqnarray}
\mathcal{L}^z&=&\sigma_1^y\sigma_2^y\sigma_4^x,\;\;  
\mathcal{L}^x=\sigma_2^y\sigma_4^z\sigma_5^z.
\end{eqnarray}
The rank of $\mathcal{X}$ in this case is four. Keeping the ordering of qubits to be unchanged, and considering the ordering for the stabilizers as $\mathcal{S}_1^c$, $\mathcal{S}_2^c$, $\mathcal{S}_3^c$, $\mathcal{S}_4^c$, and $\mathcal{L}^z$, one obtains
\begin{eqnarray}
    \mathcal{X}_{1l}&=&\begin{pmatrix}
        1 & 0 & 1 & 0 \\
        0 & 1 & 0 & 1 \\
        0 & 0 & 1 & 0 \\
        1 & 0 & 0 & 1    \end{pmatrix},\;\;\mathcal{X}_{1r}=\begin{pmatrix}
        0 & 1 & 0 & 0
    \end{pmatrix},\;
    \mathcal{Z}_{1l}=\begin{pmatrix}
        0 & 0 & 0 & 1\\
        1 & 0 & 0 & 0\\
        1 & 1 & 0 & 0\\
        0 & 1 & 1 & 0        \end{pmatrix},\;\mathcal{Z}_{1r}=\begin{pmatrix}
        0 & 0 & 1 & 1
    \end{pmatrix},\;
    \mathcal{Z}_{2l}=\begin{pmatrix}
        1\\
        1\\
        1\\
        1
        \end{pmatrix},\;\mathcal{Z}_{2r}=\begin{pmatrix}
        1
    \end{pmatrix},
\end{eqnarray}
and 
\begin{eqnarray}
    \mathcal{X}_{1l}^{-1}&=&\begin{pmatrix}
        1 & 0 & 1 & 0 \\
        1 & 1 & 1 & 1 \\
        0 & 0 & 1 & 0 \\
        1 & 0 & 1 & 1    \end{pmatrix},\;\;\mathcal{Z}_{2r}^{-1}=\begin{pmatrix}
        1
    \end{pmatrix}.
\end{eqnarray}
Substituting these in Eqs.~(\ref{eq:Uz_arbitrary})-(\ref{eq:arbitrary_stabilizer_J}),  
\begin{equation}
    R_l=I,\;\;J=\begin{pmatrix}
        1 & 0 & 1 & 0 & 0\\
        1 & 1 & 1 & 1 & 0\\
        0 & 0 & 1 & 0 & 0\\
        1 & 0 & 1 & 1 & 0\\           
        1 & 0 & 0 & 1 & 1           
    \end{pmatrix},
\end{equation}
leading to the graph generators as 
\begin{eqnarray}
g_1&=&\mathcal{S}_1^c\mathcal{S}_2^c\mathcal{S}_4^c\mathcal{L}^z,\;\;
g_2=\mathcal{S}_2^c,\;\;
g_3=\mathcal{S}_1^c\mathcal{S}_2^c\mathcal{S}_3^c\mathcal{S}_4^c,\;\;
g_4=\mathcal{S}_2^c\mathcal{S}_4^c\mathcal{L}^z,\;\;
g_5=\mathcal{L}^z,
\end{eqnarray}
and subsequently the adjacency matrix as
\begin{eqnarray}
    \Gamma=\begin{pmatrix}
        0 & 0 & 1 & 0 & 1 \\
        0 & 0 & 1 & 1 & 1 \\
        1 & 1 & 0 & 0 & 1 \\
        0 & 1 & 0 & 0 & 1 \\
        1 & 1 & 1 & 1 & 0
    \end{pmatrix}.
\end{eqnarray}
See Fig.~\ref{fig:5_1_3_perfect_code} for a schematic representation of the associated graph for the $[[5,1,3]]$ perfect code. 

\section{List of non-CSS codes}
\label{app:list_of_non_CSS}

In Tables.~\ref{tab:non_CSS_1} and \ref{tab:non_CSS_2}, we provide the set of stabilizers and logical operators for the non-CSS codes used in this paper. See Appendix.~\ref{app:5_1_3_perfect_code} for a detailed discussion on the $[[5,1,3]]$ perfect code.

\begin{table*}[]
\begin{tabular}{|p{0.4\textwidth}|p{0.55\textwidth}|}
\hline
\hspace{0.16\textwidth}$[[25,1,9]]$ \textbf{Code} & \hspace{0.23\textwidth}$[[29,1,11]]$ \textbf{Code} \\[1ex]
\multicolumn{1}{|l|}{
\begin{tabular}[c]{@{}l@{}}
\hspace{0.5cm}$\mathcal{S}_1^c = \sigma_1^x \sigma_4^y \sigma_5^y \sigma_{13}^z \sigma_{14}^y \sigma_{15}^x \sigma_{18}^x \sigma_{19}^z \sigma_{20}^y \sigma_{23}^y \sigma_{24}^x \sigma_{25}^z$ \\    
\hspace{0.5cm}$\mathcal{S}_2^c = \sigma_1^z \sigma_4^x \sigma_5^x \sigma_{13}^y \sigma_{14}^x \sigma_{15}^z \sigma_{18}^z \sigma_{19}^y \sigma_{20}^x \sigma_{23}^x \sigma_{24}^z \sigma_{25}^y$\\
\hspace{0.5cm}$\mathcal{S}_3^c = \sigma_2^x \sigma_4^y \sigma_5^z \sigma_{13}^y \sigma_{14}^x \sigma_{15}^z \sigma_{18}^z \sigma_{19}^y \sigma_{20}^x \sigma_{23}^x \sigma_{24}^z \sigma_{25}^y$\\
\hspace{0.5cm}$\mathcal{S}_4^c = \sigma_2^z \sigma_4^x \sigma_5^y \sigma_{13}^x \sigma_{14}^z \sigma_{15}^y \sigma_{18}^y \sigma_{19}^x \sigma_{20}^z \sigma_{23}^z \sigma_{24}^y \sigma_{25}^x$\\
\hspace{0.5cm}$\mathcal{S}_5^c = \sigma_3^x \sigma_4^z \sigma_5^y \sigma_{13}^y \sigma_{14}^x \sigma_{15}^z \sigma_{18}^z \sigma_{19}^y \sigma_{20}^x \sigma_{23}^x \sigma_{24}^z \sigma_{25}^y$\\
\hspace{0.5cm}$\mathcal{S}_6^c = \sigma_3^z \sigma_4^y \sigma_5^x \sigma_{13}^x \sigma_{14}^z \sigma_{15}^y \sigma_{18}^y \sigma_{19}^x \sigma_{20}^z \sigma_{23}^z \sigma_{24}^y \sigma_{25}^x$\\
\hspace{0.5cm}$\mathcal{S}_7^c = \sigma_6^x \sigma_9^y \sigma_{10}^y \sigma_{13}^y \sigma_{14}^x \sigma_{15}^z \sigma_{18}^y \sigma_{19}^x \sigma_{20}^z \sigma_{23}^y \sigma_{24}^x \sigma_{25}^z$\\
\hspace{0.5cm}$\mathcal{S}_8^c = \sigma_6^z \sigma_9^x \sigma_{10}^x \sigma_{13}^x \sigma_{14}^z \sigma_{15}^y \sigma_{18}^x \sigma_{19}^z \sigma_{20}^y \sigma_{23}^x \sigma_{24}^z \sigma_{25}^y$\\
\hspace{0.5cm}$\mathcal{S}_9^c = \sigma_7^x \sigma_9^y \sigma_{10}^z \sigma_{13}^x \sigma_{14}^z \sigma_{15}^y \sigma_{18}^x \sigma_{19}^z \sigma_{20}^y \sigma_{23}^x \sigma_{24}^z \sigma_{25}^y$\\
\hspace{0.5cm}$\mathcal{S}_{10}^c = \sigma_7^z \sigma_9^x \sigma_{10}^y \sigma_{13}^z \sigma_{14}^y \sigma_{15}^x \sigma_{18}^z \sigma_{19}^y \sigma_{20}^x \sigma_{23}^z \sigma_{24}^y \sigma_{25}^x$\\
\hspace{0.5cm}$\mathcal{S}_{11}^c = \sigma_8^x \sigma_9^z \sigma_{10}^y \sigma_{13}^x \sigma_{14}^z \sigma_{15}^y \sigma_{18}^x \sigma_{19}^z \sigma_{20}^y \sigma_{23}^x \sigma_{24}^z \sigma_{25}^y$\\
\hspace{0.5cm}$\mathcal{S}_{12}^c = \sigma_8^z \sigma_9^y \sigma_{10}^x \sigma_{13}^z \sigma_{14}^y \sigma_{15}^x \sigma_{18}^z \sigma_{19}^y \sigma_{20}^x \sigma_{23}^z \sigma_{24}^y \sigma_{25}^x$\\
\hspace{0.5cm}$\mathcal{S}_{13}^c = \sigma_{11}^x \sigma_{13}^y \sigma_{14}^z \sigma_{15}^x$\\
\hspace{0.5cm}$\mathcal{S}_{14}^c = \sigma_{11}^z \sigma_{13}^x \sigma_{14}^y \sigma_{15}^z$\\
\hspace{0.5cm}$\mathcal{S}_{15}^c = \sigma_{12}^x \sigma_{13}^x \sigma_{14}^x \sigma_{15}^x$\\
\hspace{0.5cm}$\mathcal{S}_{16}^c = \sigma_{12}^z \sigma_{13}^z \sigma_{14}^z \sigma_{15}^z$\\
\hspace{0.5cm}$\mathcal{S}_{17}^c = \sigma_{16}^x \sigma_{18}^y \sigma_{19}^z \sigma_{20}^x$\\
\hspace{0.5cm}$\mathcal{S}_{18}^c = \sigma_{16}^z \sigma_{18}^x \sigma_{19}^y \sigma_{20}^z$\\
\hspace{0.5cm}$\mathcal{S}_{19}^c = \sigma_{17}^x \sigma_{18}^x \sigma_{19}^x \sigma_{20}^x$\\
\hspace{0.5cm}$\mathcal{S}_{20}^c = \sigma_{17}^z \sigma_{18}^z \sigma_{19}^z \sigma_{20}^z$\\
\hspace{0.5cm}$\mathcal{S}_{21}^c = \sigma_{21}^x \sigma_{23}^y \sigma_{24}^z \sigma_{25}^x$\\
\hspace{0.5cm}$\mathcal{S}_{22}^c = \sigma_{21}^z \sigma_{23}^x \sigma_{24}^y \sigma_{25}^z$\\
\hspace{0.5cm}$\mathcal{S}_{23}^c = \sigma_{22}^x \sigma_{23}^x \sigma_{24}^x \sigma_{25}^x$\\
\hspace{0.5cm}$\mathcal{S}_{24}^c = \sigma_{22}^z \sigma_{23}^z \sigma_{24}^z \sigma_{25}^z$\\
\hspace{0.5cm}$\mathcal{L}^z = \sigma_1^x \sigma_4^y \sigma_5^y \sigma_7^y \sigma_8^z \sigma_9^x \sigma_{16}^x \sigma_{18}^z \sigma_{20}^z$\\
\hspace{0.5cm}$\mathcal{L}^x = \sigma_6^y \sigma_8^y \sigma_9^y \sigma_{17}^y \sigma_{19}^z \sigma_{20}^x \sigma_{22}^y \sigma_{23}^x \sigma_{25}^z$
\end{tabular}
}
&
\multicolumn{1}{l|}{
\begin{tabular}[c]{@{}l@{}}
\vspace{1pt}\\
\hspace{0.8cm}$\mathcal{S}_1^c = \sigma_1^x \sigma_{15}^x \sigma_{16}^y \sigma_{17}^z \sigma_{18}^z \sigma_{19}^x \sigma_{20}^z \sigma_{21}^z \sigma_{22}^x \sigma_{23}^x \sigma_{24}^z \sigma_{25}^z \sigma_{26}^x \sigma_{27}^z \sigma_{28}^z \sigma_{29}^y$\\
\hspace{0.8cm}$\mathcal{S}_2^c = \sigma_1^z \sigma_{15}^z \sigma_{16}^x \sigma_{17}^y \sigma_{18}^y \sigma_{19}^z \sigma_{20}^y \sigma_{21}^y \sigma_{22}^z \sigma_{23}^z \sigma_{24}^y \sigma_{25}^y \sigma_{26}^z \sigma_{27}^y \sigma_{28}^y \sigma_{29}^x$\\
\hspace{0.8cm}$\mathcal{S}_3^c = \sigma_2^x \sigma_{15}^y \sigma_{16}^y \sigma_{17}^z \sigma_{18}^y \sigma_{19}^x \sigma_{21}^y \sigma_{22}^x \sigma_{23}^z \sigma_{25}^y \sigma_{26}^x \sigma_{28}^y$\\
\hspace{0.8cm}$\mathcal{S}_4^c = \sigma_2^z \sigma_{15}^x \sigma_{16}^x \sigma_{17}^y \sigma_{18}^x \sigma_{19}^z \sigma_{21}^x \sigma_{22}^z \sigma_{23}^y \sigma_{25}^x \sigma_{26}^z \sigma_{28}^x$\\
\hspace{0.8cm}$\mathcal{S}_5^c = \sigma_3^x \sigma_{16}^y \sigma_{17}^y \sigma_{18}^z \sigma_{19}^y \sigma_{20}^x \sigma_{22}^y \sigma_{23}^x \sigma_{24}^z \sigma_{26}^y \sigma_{27}^x \sigma_{29}^y$\\
\hspace{0.8cm}$\mathcal{S}_6^c = \sigma_3^z \sigma_{16}^x \sigma_{17}^x \sigma_{18}^y \sigma_{19}^x \sigma_{20}^z \sigma_{22}^x \sigma_{23}^z \sigma_{24}^y \sigma_{26}^x \sigma_{27}^z \sigma_{29}^x$\\
\hspace{0.8cm}$\mathcal{S}_7^c = \sigma_4^x \sigma_{15}^y \sigma_{16}^z \sigma_{17}^z \sigma_{18}^z \sigma_{19}^x \sigma_{20}^z \sigma_{22}^y \sigma_{25}^y \sigma_{26}^y \sigma_{27}^z \sigma_{29}^z$\\
\hspace{0.8cm}$\mathcal{S}_8^c = \sigma_4^z \sigma_{15}^x \sigma_{16}^y \sigma_{17}^y \sigma_{18}^y \sigma_{19}^z \sigma_{20}^y \sigma_{22}^x \sigma_{25}^x \sigma_{26}^x \sigma_{27}^y \sigma_{29}^y$\\
\hspace{0.8cm}$\mathcal{S}_9^c = \sigma_5^x \sigma_{15}^z \sigma_{16}^z \sigma_{17}^x \sigma_{18}^x \sigma_{20}^z \sigma_{21}^x \sigma_{22}^z \sigma_{23}^x \sigma_{24}^y \sigma_{25}^y \sigma_{26}^x \sigma_{28}^x \sigma_{29}^x$\\
\hspace{0.8cm}$\mathcal{S}_{10}^c = \sigma_5^z \sigma_{15}^y \sigma_{16}^y \sigma_{17}^z \sigma_{18}^z \sigma_{20}^y \sigma_{21}^z \sigma_{22}^y \sigma_{23}^z \sigma_{24}^x \sigma_{25}^x \sigma_{26}^z \sigma_{28}^z \sigma_{29}^z$\\
\hspace{0.8cm}$\mathcal{S}_{11}^c = \sigma_6^x \sigma_{15}^x \sigma_{16}^x \sigma_{18}^y \sigma_{20}^z \sigma_{23}^y \sigma_{24}^y \sigma_{25}^x \sigma_{26}^z \sigma_{27}^y \sigma_{28}^z \sigma_{29}^z$\\
\hspace{0.8cm}$\mathcal{S}_{12}^c = \sigma_6^z \sigma_{15}^z \sigma_{16}^z \sigma_{18}^x \sigma_{20}^y \sigma_{23}^x \sigma_{24}^x \sigma_{25}^z \sigma_{26}^y \sigma_{27}^x \sigma_{28}^y \sigma_{29}^y$\\
\hspace{0.8cm}$\mathcal{S}_{13}^c = \sigma_7^x \sigma_{15}^z \sigma_{17}^z \sigma_{18}^y \sigma_{19}^x \sigma_{20}^y \sigma_{21}^x \sigma_{22}^z \sigma_{23}^z \sigma_{26}^y \sigma_{27}^x \sigma_{29}^y$\\
\hspace{0.8cm}$\mathcal{S}_{14}^c = \sigma_7^z \sigma_{15}^y \sigma_{17}^y \sigma_{18}^x \sigma_{19}^z \sigma_{20}^x \sigma_{21}^z \sigma_{22}^y \sigma_{23}^y \sigma_{26}^x \sigma_{27}^z \sigma_{29}^x$\\
\hspace{0.8cm}$\mathcal{S}_{15}^c = \sigma_8^x \sigma_{15}^y \sigma_{17}^x \sigma_{18}^y \sigma_{21}^z \sigma_{22}^z \sigma_{23}^x \sigma_{24}^y \sigma_{25}^x \sigma_{26}^y \sigma_{27}^z \sigma_{29}^z$\\
\hspace{0.8cm}$\mathcal{S}_{16}^c = \sigma_8^z \sigma_{15}^x \sigma_{17}^z \sigma_{18}^x \sigma_{21}^y \sigma_{22}^y \sigma_{23}^z \sigma_{24}^x \sigma_{25}^z \sigma_{26}^x \sigma_{27}^y \sigma_{29}^y$\\
\hspace{0.8cm}$\mathcal{S}_{17}^c = \sigma_9^x \sigma_{15}^z \sigma_{16}^z \sigma_{17}^y \sigma_{18}^z \sigma_{19}^x \sigma_{20}^y \sigma_{21}^y \sigma_{24}^z \sigma_{26}^y \sigma_{28}^x \sigma_{29}^x$\\
\hspace{0.8cm}$\mathcal{S}_{18}^c = \sigma_9^z \sigma_{15}^y \sigma_{16}^y \sigma_{17}^x \sigma_{18}^y \sigma_{19}^z \sigma_{20}^x \sigma_{21}^x \sigma_{24}^y \sigma_{26}^x \sigma_{28}^z \sigma_{29}^z$\\
\hspace{0.8cm}$\mathcal{S}_{19}^c = \sigma_{10}^x \sigma_{15}^x \sigma_{16}^x \sigma_{18}^x \sigma_{19}^y \sigma_{20}^y \sigma_{21}^x \sigma_{22}^z \sigma_{23}^x \sigma_{24}^z \sigma_{26}^x \sigma_{27}^x \sigma_{28}^z \sigma_{29}^z$\\
\hspace{0.8cm}$\mathcal{S}_{20}^c = \sigma_{10}^z \sigma_{15}^z \sigma_{16}^z \sigma_{18}^z \sigma_{19}^x \sigma_{20}^x \sigma_{21}^z \sigma_{22}^y \sigma_{23}^z \sigma_{24}^y \sigma_{26}^z \sigma_{27}^z \sigma_{28}^y \sigma_{29}^y$\\
\hspace{0.8cm}$\mathcal{S}_{21}^c = \sigma_{11}^x \sigma_{15}^z \sigma_{17}^z \sigma_{18}^y \sigma_{19}^y \sigma_{22}^y \sigma_{24}^z \sigma_{25}^x \sigma_{26}^z \sigma_{27}^z \sigma_{28}^z \sigma_{29}^y$\\
\hspace{0.8cm}$\mathcal{S}_{22}^c = \sigma_{11}^z \sigma_{15}^y \sigma_{17}^y \sigma_{18}^x \sigma_{19}^x \sigma_{22}^x \sigma_{24}^y \sigma_{25}^z \sigma_{26}^y \sigma_{27}^y \sigma_{28}^y \sigma_{29}^x$\\
\hspace{0.8cm}$\mathcal{S}_{23}^c = \sigma_{12}^x \sigma_{15}^y \sigma_{17}^x \sigma_{18}^y \sigma_{20}^z \sigma_{21}^x \sigma_{22}^y \sigma_{24}^x \sigma_{25}^y \sigma_{26}^z \sigma_{27}^y \sigma_{28}^y$\\
\hspace{0.8cm}$\mathcal{S}_{24}^c = \sigma_{12}^z \sigma_{15}^x \sigma_{17}^z \sigma_{18}^x \sigma_{20}^y \sigma_{21}^z \sigma_{22}^x \sigma_{24}^z \sigma_{25}^x \sigma_{26}^y \sigma_{27}^x \sigma_{28}^x$\\
\hspace{0.8cm}$\mathcal{S}_{25}^c = \sigma_{13}^x \sigma_{16}^y \sigma_{18}^x \sigma_{19}^y \sigma_{21}^z \sigma_{22}^x \sigma_{23}^y \sigma_{25}^x \sigma_{26}^y \sigma_{27}^z \sigma_{28}^y \sigma_{29}^y$\\
\hspace{0.8cm}$\mathcal{S}_{26}^c = \sigma_{13}^z \sigma_{16}^x \sigma_{18}^z \sigma_{19}^x \sigma_{21}^y \sigma_{22}^z \sigma_{23}^x \sigma_{25}^z \sigma_{26}^x \sigma_{27}^y \sigma_{28}^x \sigma_{29}^x$\\
\hspace{0.8cm}$\mathcal{S}_{27}^c = \sigma_{14}^x \sigma_{15}^y \sigma_{16}^z \sigma_{17}^z \sigma_{18}^x \sigma_{19}^z \sigma_{20}^z \sigma_{21}^x \sigma_{22}^x \sigma_{23}^z \sigma_{24}^z \sigma_{25}^x \sigma_{26}^z \sigma_{27}^z \sigma_{28}^y \sigma_{29}^x$\\
\hspace{0.8cm}$\mathcal{S}_{28}^c = \sigma_{14}^z \sigma_{15}^x \sigma_{16}^y \sigma_{17}^y \sigma_{18}^z \sigma_{19}^y \sigma_{20}^y \sigma_{21}^z \sigma_{22}^z \sigma_{23}^y \sigma_{24}^y \sigma_{25}^z \sigma_{26}^y \sigma_{27}^y \sigma_{28}^x \sigma_{29}^z$\\
\hspace{0.8cm}$\mathcal{L}^z = \sigma_4^z \sigma_{10}^x \sigma_{12}^x \sigma_{13}^y \sigma_{14}^y \sigma_{15}^z \sigma_{17}^y \sigma_{22}^x \sigma_{24}^x \sigma_{27}^x \sigma_{29}^y$\\
\hspace{0.8cm}$\mathcal{L}^x = \sigma_5^z \sigma_8^x \sigma_9^y \sigma_{12}^x \sigma_{15}^y \sigma_{18}^x \sigma_{20}^z \sigma_{24}^y \sigma_{26}^y \sigma_{27}^x \sigma_{29}^z$\\
\vspace{1pt}
\end{tabular}
}
\\ \hline

\end{tabular}

\caption{Stabilizers for the $[[25,1,9]]$ and $[[29,1,11]]$ codes belonging to the non-CSS family.}\label{tab:non_CSS_2}
\end{table*}

\section{The \texorpdfstring{$[[7,1,3]]$} code as an example of CSS codes}
\label{app:7_1_3_color_code}

The $[[7,1,3]]$ code is the smallest of the CSS code family. The stabilizers of the $[[7,1,3]]$ code are given by 
\begin{eqnarray}
\mathcal{S}_1^{c,x(z)}&=&\sigma_1^{x(z)}\sigma_2^{x(z)}\sigma_3^{x(z)}\sigma_4^{x(z)},\;\;
\mathcal{S}_2^{c,x(z)}=\sigma_2^{x(z)}\sigma_3^{x(z)}\sigma_5^{x(z)}\sigma_6^{x(z)},\;\;
\mathcal{S}_3^{c,x(z)}=\sigma_3^{x(z)}\sigma_4^{x(z)}\sigma_6^{x(z)}\sigma_7^{x(z)},    
\end{eqnarray}
while the logical operators are 
\begin{eqnarray}
\mathcal{L}^z=\sigma_5^z\sigma_6^z\sigma_7^z,\;\;
\mathcal{L}^x=\sigma_1^x\sigma_2^x\sigma_5^x.
\end{eqnarray}
The rank of $\mathcal{X}$, in this case, is three.  Considering the ordering of nodes as $1,5$, and $7$ ($2,3,4$, and $6$) in the set of the left (right) nodes, and keeping the ordering of stabilizers to be $\mathcal{S}_1^{c,x}$, $\mathcal{S}_2^{c,x}$, $\mathcal{S}_3^{c,x}$, $\mathcal{S}_1^{c,z}$, $\mathcal{S}_2^{c,z}$, $\mathcal{S}_3^{c,z}$, and $\mathcal{L}^z$, one obtains 
\begin{eqnarray}
    \mathcal{X}_{1l}&=&\begin{pmatrix}
        1 & 0 & 0 \\
        0 & 1 & 0 \\
        0 & 0 & 1 
    \end{pmatrix},\;\;\mathcal{X}_{1r}=\begin{pmatrix}
        1 & 1 & 0 \\
        1 & 1 & 1 \\
        1 & 0 & 1 \\
        0 & 1 & 1
    \end{pmatrix},\;\;
    \mathcal{Z}_{2l}=\begin{pmatrix}
        1 & 0 & 0 & 0\\
        0 & 1 & 0 & 1\\
        0 & 0 & 1 & 1    \end{pmatrix},\;\;\mathcal{Z}_{2r}=\begin{pmatrix}
        1 & 1 & 0 & 0\\
        1 & 1 & 1 & 0\\
        1 & 0 & 1 & 0\\
        0 & 1 & 1 & 1
    \end{pmatrix},
\end{eqnarray}
resulting in  
\begin{eqnarray}
    \mathcal{X}_{1l}^{-1}=\begin{pmatrix}
        1 & 0 & 0 \\
        0 & 1 & 0 \\
        0 & 0 & 1     \end{pmatrix},\;\;\mathcal{Z}_{2r}^{-1}=\begin{pmatrix}
        1 & 1 & 1 & 0\\
        0 & 1 & 1 & 0\\
        1 & 1 & 0 & 0\\
        1 & 0 & 1 & 1
    \end{pmatrix},
\end{eqnarray}
and subsequently 
\begin{eqnarray}
    J=\begin{pmatrix}
        1 & 0 & 0 & 0 & 0 & 0 & 0 \\
        0 & 1 & 0 & 0 & 0 & 0 & 0 \\
        0 & 0 & 1 & 0 & 0 & 0 & 0 \\
        0 & 0 & 0 & 1 & 1 & 1 & 0\\
        0 & 0 & 0 & 0 & 1 & 1 & 0\\
        0 & 0 & 0 & 1 & 1 & 0 & 0\\
        0 & 0 & 0 & 1 & 0 & 1 & 1
    \end{pmatrix}.
\end{eqnarray}
This leads to the graph generators,
\begin{eqnarray}
g_1&=&\mathcal{S}_1^{c,x},\;\;
g_2=\mathcal{S}_1^{c,z}\mathcal{S}_3^{c,z}\mathcal{L}^z,\;\;
g_3=\mathcal{S}_1^{c,z}\mathcal{S}_2^{c,z}\mathcal{S}_3^{c,z},
g_4=\mathcal{S}_1^{c,z}\mathcal{S}_2^{c,z}\mathcal{L}^z,\;\;
g_5=\mathcal{S}_2^{c,x},\;\;
g_6=\mathcal{L}^{c,z},\;\;
g_7=\mathcal{S}_3^{c,x}, 
\end{eqnarray}
and subsequently the adjacency matrix (see Fig.~\ref{fig:7_1_3_CSS} for a schematic representation),
\begin{eqnarray}
    \Gamma=\begin{pmatrix}
        0 & 0 & 0 & 1 & 1 & 1 & 0 \\
        0 & 0 & 0 & 1 & 1 & 0 & 1 \\
        0 & 0 & 0 & 0 & 1 & 1 & 1 \\
        1 & 1 & 0 & 0 & 0 & 0 & 0\\
        1 & 1 & 1 & 0 & 0 & 0 & 0\\
        1 & 0 & 1 & 0 & 0 & 0 & 0\\
        0 & 1 & 1 & 0 & 0 & 0 & 0
    \end{pmatrix}.
\end{eqnarray}

\twocolumngrid 

\bibliography{ref}

\end{document}